\documentclass[conference]{IEEEtran}

\usepackage{cite}
\usepackage{amsmath,amssymb,amsfonts}
\usepackage{algorithmic}
\usepackage{graphicx}
\usepackage{textcomp}
\usepackage{xcolor}
\def\BibTeX{{\rm B\kern-.05em{\sc i\kern-.025em b}\kern-.08em
    T\kern-.1667em\lower.7ex\hbox{E}\kern-.125emX}}

\usepackage[T1]{fontenc}

\usepackage{scalerel}
\usepackage{xfrac} % split-level fractions

\usepackage{siunitx}
\DeclareSIPrefix\micro{\text{\textmu}}{-3}

\usepackage{tikz}
\usepackage{pgfplots}
\pgfplotsset{compat=newest}
\usetikzlibrary{positioning,3d,calc,intersections,svg.path,pgfplots.external,arrows.meta,backgrounds,fit}
\usepgfplotslibrary{statistics}
\usepackage{array}	% include commands for table column specifications, e.g. @{}, >{$} etc.
\usepackage{booktabs}	% format tables nicely
\usepackage{dcolumn}	% define custom column syles
\usepackage{multirow}
\usepackage{multicol}
\usepackage{accents}	% additional math accents, e.g. \underaccent

\usepackage{marginnote}
\newcommand{\TODO}[1]{}

\newcolumntype{H}{>{\scriptsize}c}

\usepackage[labelformat=simple,labelfont=bf,font=footnotesize,singlelinecheck=true]{subcaption}	% Bilder mit (a), (b) usw. bezeichnen und referenzieren

\DeclareCaptionLabelSeparator{periodspace}{.\quad}
\definecolor{orcidlogocol}{HTML}{A6CE39}
\tikzset{
  orcidlogo/.pic={
    \fill[orcidlogocol] svg{M256,128c0,70.7-57.3,128-128,128C57.3,256,0,198.7,0,128C0,57.3,57.3,0,128,0C198.7,0,256,57.3,256,128z};
    \fill[white] svg{M86.3,186.2H70.9V79.1h15.4v48.4V186.2z}
                 svg{M108.9,79.1h41.6c39.6,0,57,28.3,57,53.6c0,27.5-21.5,53.6-56.8,53.6h-41.8V79.1z M124.3,172.4h24.5c34.9,0,42.9-26.5,42.9-39.7c0-21.5-13.7-39.7-43.7-39.7h-23.7V172.4z}
                 svg{M88.7,56.8c0,5.5-4.5,10.1-10.1,10.1c-5.6,0-10.1-4.6-10.1-10.1c0-5.6,4.5-10.1,10.1-10.1C84.2,46.7,88.7,51.3,88.7,56.8z};
  }
}

\newcommand\orcidlink[1]{\href{https://orcid.org/#1}{\mbox{\scalerel*{
\begin{tikzpicture}[yscale=-1,transform shape]
\pic{orcidlogo};
\end{tikzpicture}
}{|}}}}

\usepackage{hyperref}

\hypersetup{%
  pdfauthor={M. F. Schiffner},%
  pdftitle={Laplace-Domain Beamforming for Ultrafast Plane-Wave Imaging},%
  pdfsubject={Lehrstuhl für Medizintechnik, Ruhr-Universität Bochum},%
  pdfkeywords={%
    ultrafast ultrasound imaging, %
    Fourier-domain beamforming, %
    Fourier diffraction theorem, %
    Laplace diffraction theorem, %
    diffraction tomography, %
    inverse scattering%
  },%
  colorlinks=false,%
  hidelinks,
  linktocpage=true,%
  bookmarksopenlevel=1%
}

\usepackage{breakurl} % advanced line breaks for URLs

\usepackage{acronym} % definition of acronyms

\usepackage{cleveref}	  			% package for clever references

\crefname{chapter}{Chapter}{Chapters}		% in the middle of sentence
\Crefname{chapter}{Chapter}{Chapters}		% beginning of sentence
\crefname{appendix}{Appendix}{Appendices}
\Crefname{appendix}{Appendix}{Appendices}
\crefname{section}{Sect.}{Sections}
\Crefname{section}{Section}{Sections}
\crefname{subsection}{Sect.}{Sections}
\Crefname{subsection}{Section}{Sections}
\crefname{subsubsection}{Sect.}{Sections}
\Crefname{subsubsection}{Section}{Sections}

\crefname{figure}{Fig.}{Figs.}
\Crefname{figure}{Figure}{Figures}

\crefname{table}{Table}{Tables}
\Crefname{table}{Table}{Tables}

\newcommand{\name}[1]{#1}

\definecolor{RUBBLUE}{cmyk}{1.000,0.500,0.000,0.600}	% PANTONE 540 (CMYK 100%, 50%, 0%, 60%) RAL 5003 (SAPHIRBLAU)
\definecolor{RUBGREEN}{cmyk}{0.500,0.000,1.000,0.000}	% PANTONE 376 (CMYK 50%, 0%, 100%, 0%)  RAL 6018 (GELBGRÜN)
\definecolor{RUBGRAY}{cmyk}{0.030,0.030,0.030,0.100}	% PANTONE Cool Grey 2 (CMYK 3%, 3%, 3%, 10%) RAL 7035 (LICHTGRAU)

\definecolor{gray}{gray}{0.5}
\definecolor{gray70}{gray}{0.3}
\definecolor{gray60}{gray}{0.4}
\definecolor{gray40}{gray}{0.6}
\definecolor{gray30}{gray}{0.7}
\definecolor{gray20}{gray}{0.8}
\definecolor{gray10}{gray}{0.9}
\definecolor{gray05}{gray}{0.95}

\newcommand{\setsymbol}[1]{\ensuremath{\mathbb{#1}}}

\newcommand{\R}{\setsymbol{R}}

\newcommand{\C}{\setsymbol{C}}

\newcommand{\dsetcons}[2]%
{%
\ensuremath%
 {%
  \ifthenelse{\equal{#2}{0}}{[ #1 ]}{}%
  \ifthenelse{\equal{#2}{1}}{\bigl[ #1 \bigr]}{}%
  \ifthenelse{\equal{#2}{2}}{\Bigl[ #1 \Bigr]}{}%
  \ifthenelse{\equal{#2}{3}}{\biggl[ #1 \biggr]}{}%
  \ifthenelse{\equal{#2}{4}}{\Biggl[ #1 \Biggr]}{}%
 }%
}

\newcommand{\dsetconsnonneg}[2]%
{%
\ensuremath%
 {%
  \ifthenelse{\equal{#2}{0}}{[ #1 ]_{0}}{}%
  \ifthenelse{\equal{#2}{1}}{\bigl[ #1 \bigr]_{0}}{}%
  \ifthenelse{\equal{#2}{2}}{\Bigl[ #1 \Bigr]_{0}}{}%
  \ifthenelse{\equal{#2}{3}}{\biggl[ #1 \biggr]_{0}}{}%
  \ifthenelse{\equal{#2}{4}}{\Biggl[ #1 \Biggr]_{0}}{}%
 }%
}

\newcommand{\disksymbol}{\ensuremath{\setsymbol{D}}}
\newcommand{\opendisk}[1]{\ensuremath{\disksymbol_{#1}}}

\newcommand{\vect}[1]{\ensuremath{\mathbf{#1}}}
\newcommand{\vectsym}[1]{\ensuremath{\boldsymbol{#1}}}

\newcommand{\uvectcompsymbol}{\ensuremath{e}}

\newcommand{\uvectsymbol}{\ensuremath{\vect{\uvectcompsymbol}}}
\newcommand{\uvect}[1]{\ensuremath{\uvectsymbol_{#1}}}

\newcommand{\mat}[1]{\ensuremath{\mathbf{#1}}}
\newcommand{\transsymbol}{\ensuremath{\mathrm{T}}}
\newcommand{\trans}[1]{\ensuremath{#1^{\transsymbol}}}

\newcommand{\diagsymbol}{\ensuremath{\mathrm{diag}}}

\newcommand{\ddiag}[2]%
{%
 \ensuremath
 {%
  \ifthenelse{\equal{#2}{0}}{\diagsymbol\{ #1 \}}{}%
  \ifthenelse{\equal{#2}{1}}{\diagsymbol\bigl\{ #1 \bigr\}}{}%
  \ifthenelse{\equal{#2}{2}}{\diagsymbol\Bigl\{ #1 \Bigr\}}{}%
  \ifthenelse{\equal{#2}{3}}{\diagsymbol\biggl\{ #1\biggr \}}{}%
  \ifthenelse{\equal{#2}{4}}{\diagsymbol\Biggl\{ #1\Biggr \}}{}%
 }%
}

\newcommand{\dabs}[2]
{%
 \ensuremath%
 {
  \ifthenelse{\equal{#2}{0}}{\lvert#1\rvert}{}%
  \ifthenelse{\equal{#2}{1}}{\bigl\lvert#1\bigr\rvert}{}%
  \ifthenelse{\equal{#2}{2}}{\Bigl\lvert#1\Bigr\rvert}{}%
  \ifthenelse{\equal{#2}{3}}{\biggl\lvert#1\biggr\rvert}{}%
  \ifthenelse{\equal{#2}{4}}{\Biggl\lvert#1\Biggr\rvert}{}%
 }%
}
\newcommand{\tabs}[1]{\ensuremath{\lvert#1\rvert}}

\newcommand{\inprod}[2]{\ensuremath{\left\langle #1, #2 \right\rangle}}

\newcommand{\dinprod}[3]{%
\ensuremath%
 {%
  \ifthenelse{\equal{#3}{0}}{\langle #1, #2 \rangle}{}%
  \ifthenelse{\equal{#3}{1}}{\bigl\langle #1, #2 \bigr\rangle}{}%
  \ifthenelse{\equal{#3}{2}}{\Bigl\langle #1, #2 \Bigr\rangle}{}%
  \ifthenelse{\equal{#3}{3}}{\biggl\langle #1, #2 \biggr\rangle}{}%
  \ifthenelse{\equal{#3}{4}}{\Biggl\langle #1, #2 \Biggr\rangle}{}%
 }%
}
\newcommand{\tinprod}[2]{\ensuremath{\langle #1, #2 \rangle}}

\newcommand{\dinprodr}[3]{%
\ensuremath%
 {%
  \ifthenelse{\equal{#3}{0}}{\langle #1, #2 \rangle_{\text{r}}}{}%
  \ifthenelse{\equal{#3}{1}}{\bigl\langle #1, #2 \bigr\rangle_{\text{r}}}{}%
  \ifthenelse{\equal{#3}{2}}{\Bigl\langle #1, #2 \Bigr\rangle_{\text{r}}}{}%
  \ifthenelse{\equal{#3}{3}}{\biggl\langle #1, #2 \biggr\rangle_{\text{r}}}{}%
  \ifthenelse{\equal{#3}{4}}{\Biggl\langle #1, #2 \Biggr\rangle_{\text{r}}}{}%
 }%
}
\newcommand{\norm}[2]{\ensuremath{\left\lVert#1\right\rVert_{#2}}}
\newcommand{\dnorm}[3]%
{%
 \ensuremath%
 {%
  \ifthenelse{\equal{#3}{0}}{\lVert#1\rVert_{#2}}{}%
  \ifthenelse{\equal{#3}{1}}{\bigl\lVert#1\bigr\rVert_{#2}}{}%
  \ifthenelse{\equal{#3}{2}}{\Bigl\lVert#1\Bigr\rVert_{#2}}{}%
  \ifthenelse{\equal{#3}{3}}{\biggl\lVert#1\biggr\rVert_{#2}}{}%
  \ifthenelse{\equal{#3}{4}}{\Biggl\lVert#1\Biggr\rVert_{#2}}{}%
 }%
}
\newcommand{\tnorm}[2]{\ensuremath{\lVert#1\rVert_{#2}}}

\newcommand{\sinc}{\ensuremath{\text{sinc}}}

\newcommand{\sinhc}{\ensuremath{\text{sinhc}}}

\newcommand{\dsign}[2]
{%
 \ensuremath%
 {
  \ifthenelse{\equal{#2}{0}}{\ensuremath{\text{sgn}}(#1)}{}%
  \ifthenelse{\equal{#2}{1}}{\ensuremath{\text{sgn}}\bigl(#1\bigr)}{}%
  \ifthenelse{\equal{#2}{2}}{\ensuremath{\text{sgn}}\Bigl(#1\Bigr)}{}%
  \ifthenelse{\equal{#2}{3}}{\ensuremath{\text{sgn}}\biggl(#1\biggr)}{}%
  \ifthenelse{\equal{#2}{4}}{\ensuremath{\text{sgn}}\Biggl(#1\Biggr)}{}%
 }%
}

\newcommand{\munderbar}[1]{\ensuremath{\underaccent{\bar}{#1}}}

\newcommand{\term}[1]{``#1''}

\newcommand{\dreal}[2]
{%
 \ensuremath%
 {
  \mathrm{Re}%
  \ifthenelse{\equal{#2}{0}}{\{#1\}}{}%
  \ifthenelse{\equal{#2}{1}}{\bigl\{#1\bigr\}}{}%
  \ifthenelse{\equal{#2}{2}}{\Bigl\{#1\Bigr\}}{}%
  \ifthenelse{\equal{#2}{3}}{\biggl\{#1\biggr\}}{}%
  \ifthenelse{\equal{#2}{4}}{\Biggl\{#1\Biggr\}}{}%
 }%
}
\newcommand{\treal}[1]{\ensuremath{\mathrm{Re}\{#1\}}}

\newcommand{\imag}[1]{\ensuremath{\mathrm{Im}\left\{#1\right\}}}
\newcommand{\dimag}[2]
{%
 \ensuremath%
 {
  \mathrm{Im}%
  \ifthenelse{\equal{#2}{0}}{\{#1\}}{}%
  \ifthenelse{\equal{#2}{1}}{\bigl\{#1\bigr\}}{}%
  \ifthenelse{\equal{#2}{2}}{\Bigl\{#1\Bigr\}}{}%
  \ifthenelse{\equal{#2}{3}}{\biggl\{#1\biggr\}}{}%
  \ifthenelse{\equal{#2}{4}}{\Biggl\{#1\Biggr\}}{}%
 }%
}
\newcommand{\timag}[1]{\ensuremath{\mathrm{Im}\{#1\}}}



\acrodef{AAPM}{American Association of Physicists in Medicine}		%001
\acrodef{ACB}{asyn\-chro\-nous compressed beamformer}			%002
\acrodef{ACA}{adaptive cross approximation}				%003
\acrodef{ACR}{American College of Radiology}				%004
\acrodef{ADC}{analog-to-digital conversion}				%005
\acrodef{ADMM}{alternating direction method of multipliers}             %006
\acrodef{AIC}{analog-to-information conversion}				%007
\acrodef{BIBO}{bounded-input, bounded-output}				%008
\acrodef{BIM}{\name{Born} iterative method}				%009
\acrodef{BP}{basis pursuit}						%010

\acrodef{BPDN}{basis pursuit denoise}					%011
\acrodef{BVP}{boundary value problem}					%012
\acrodef{CAR}{Canadian Association of Radiologists}			%013
\acrodef{CDF}{cumulative distribution function}				%014
\acrodef{CGP}{conjugate gradient pursuit}				%015
\acrodef{CNR}{contrast-to-noise ratio}					%016
\acrodef{CoSaMP}{compressive sampling matching pursuit}			%017
\acrodef{CPU}{central processing unit}					%018
\acrodef{CPV}{\name{Cauchy} principal value}				%019
\acrodef{CPWC}{coherent plane-wave compounding}				%020

\acrodef{CS}{compressed sensing}					%021
\acrodef{CSI}{contrast source inversion}				%022
\acrodef{CT}{x-ray computed tomography}					%023
\acrodef{CTTE}[CT-TE]{continuous-time ternary encoding}			%024
\acrodef{CW}{cylindrical wave}						%025
\acrodef{DAC}{digital-to-analog conversion}				%026
\acrodef{DAIC}{diagnostically acceptable irreversible compression}	%027
\acrodef{DAS}{delay-and-sum}						%028
\acrodef{DBIM}{distorted \name{Born} iterative method}			%029
\acrodef{DCT}{discrete cosine transform}				%030

\acrodef{DFT}{discrete \name{Fourier} transform}			%031
\acrodef{DICOM}{Digital Imaging and Communications in Medicine}		%032
\acrodef{DNA}{deoxyribonucleic acid}					%033
\acrodef{DOF}{depth of field}                                           %034.s
\acrodefplural{DOF}{depths of field}                                    %034.p
\acrodef{DRG}{German \name{Röntgen} Society}				%035
\acrodef{DRIM}{distorted \name{Rytov} iterative method}			%036
\acrodef{DRS}{digital random sampler}					%037
\acrodef{DSFT}{discrete space \name{Fourier} transform}			%038
\acrodef{DWT}{discrete wavelet transform}				%039
\acrodef{ERM}{exploding reflector model}				%040

\acrodef{ESR}{European Society of Radiology}				%041
\acrodef{EU}{European Union}						%042
\acrodef{FBP}{filtered backpropagation}					%043
\acrodef{FDA}{US Food and Drug Administration}				%044 %U.S. Food and Drug Administration
\acrodef{FDBF}{frequency domain beamforming}				%045
\acrodef{FDT}{\name{Fourier} diffraction theorem}			%046
\acrodef{FEHM}{full extent at half maximum}				%047.s
\acrodefplural{FEHM}{full extents at half maximum}			%047.p
\acrodef{FFT}{fast \name{Fourier} transform}				%048
\acrodef{FIR}{finite impulse response}					%049
\acrodef{FISTA}{fast iterative shrinkage-thresholding algorithm}        %050

\acrodef{FLOP}{floating point operation}				%051
\acrodef{FMM}{fast multipole method}					%052
\acrodef{FOCUS}{fast object-oriented C++ ultrasound simulator}		%053
\acrodef{FOV}{field of view}						%054.s
\acrodefplural{FOV}{fields of view}					%054.p
\acrodef{FPDE}{fractional partial differential equation}		%055
\acrodef{FPS}{frames per second}					%056
\acrodef{FRI}{finite rate of innovation}				%057
\acrodef{FWHM}{full width at half maximum}				%058.s
\acrodefplural{FWHM}{full widths at half maximum}			%058.p
\acrodef{FWT}{fast wavelet transform}					%059
\acrodef{GCNR}[gCNR]{generalized contrast-to-noise ratio}		%060

\acrodef{GPU}{graphics processing unit}					%061
\acrodef{GWN}{Gaussian white noise}					%062
\acrodef{HSV}{hue, saturation, value}					%063
\acrodef{HTP}{hard thresholding pursuit}				%064
\acrodef{IHT}{iterative hard thresholding}				%065
\acrodef{IID}[i.i.d.]{independent and identically distributed}		%066
\acrodef{IQ}{in-phase and quadrature}					%067
\acrodef{IQR}{interquartile range}					%068
\acrodef{IRLS}{iteratively reweighted least squares}			%069
\acrodef{ISP}{inverse scattering problem}				%070

\acrodef{ISR}{integrated sidelobe ratio}				%071
\acrodef{JPEG}{Joint Photographic Experts Group}			%072
\acrodef{JSM}{joint sparsity model}					%073
\acrodef{KK}{\name{Kramers}-\name{Kronig}}                              %074
\acrodef{LASSO}{least absolute shrinkage and selection operator}	%075
\acrodef{LDT}{\name{Laplace} diffraction theorem}			%076
\acrodef{LPIPS}{learned perceptual image patch similarity}		%077
\acrodef{LS}{\name{Lippmann}-\name{Schwinger}}				%078
\acrodef{LSB}{least significant bit}					%079
\acrodef{LTI}{linear time-invariant}                                    %080

\acrodef{MAP}{maximum \emph{a posteriori}}				%081
\acrodef{MCMC}{\name{Markov} chain Monte Carlo}				%082
\acrodef{MIMO}{multiple-input, multiple-output}				%083
\acrodef{MIRF}{material impulse response function}			%084
\acrodef{MLFMA}{multilevel fast multipole algorithm}			%085
\acrodef{MMV}{multiple measurement vectors}				%086
\acrodef{MOM}[MoM]{method of moments}					%087
\acrodef{MP}{matching pursuit}						%088
\acrodef{MP3}{MPEG Audio Layer III}					%089
\acrodef{MRI}{magnetic resonance imaging}				%090

\acrodef{MTF}{material transfer function}				%091
\acrodef{MV}{minimum variance}						%092
\acrodef{MVDR}{minimum-variance distortionless response}		%093
\acrodef{MWC}{modulated wideband converter}				%094
\acrodef{NESTA}{\name{Nesterov}'s algorithm}				%095
\acrodef{NSP}{null space property}					%096
\acrodef{NNZC}{number of nonzero components}				%097
\acrodef{OMP}{orthogonal matching pursuit}				%098
\acrodef{PAM}{pulse-amplitude modulation}				%099
\acrodef{PAT}{photoacoustic tomography}					%100

\acrodef{PC}{personal computer}						%101
\acrodef{PCIe}{peripheral component interconnect express}               %102
\acrodef{PDE}{partial differential equation}				%103
\acrodef{PDF}{probability density function}				%104
\acrodef{PET}{positron emission tomography}				%105
\acrodef{PICMUS}{Plane-Wave Imaging Challenge in Medical Ultrasound}	%106
\acrodef{PSF}{point spread function}					%107
\acrodef{PSL}{peak sidelobe level}					%108
\acrodef{PSNR}{peak signal-to-noise ratio}				%109
\acrodef{PW}{plane wave}						%110

\acrodef{QCW}{quasi-cylindrical wave}					%111
\acrodef{QPW}{quasi-plane wave}						%112
\acrodef{QSW}{quasi-spherical wave}					%113
\acrodef{QQ}{quantile-quantile}						%114
\acrodef{RA}{randomly-apodized}						%115
\acrodef{RAD}{randomly-apodized and randomly-delayed}			%116
\acrodef{RADAR}[RADAR]{radio detection and ranging}			%117
\acrodef{RAM}{random-access memory}					%118
\acrodef{RCR}{Royal College of Radiologists}				%119
\acrodef{RD}{randomly-delayed}						%120

\acrodef{RF}{radio frequency}						%121
\acrodef{RIC}{restricted isometry constant}				%122
\acrodef{RIP}{restricted isometry property}				%123
\acrodef{RMSE}{root mean-squared error}					%124
\acrodef{ROI}{region of interest}					%125.s
\acrodefplural{ROI}{regions of interest}				%125.p
\acrodef{ROMP}{regularized \ac{OMP}}					%126
\acrodef{RS}{\name{Rayleigh}-\name{Sommerfeld}}                         %127
\acrodef{RW}{randomly-weighted}                                         %128
\acrodef{RWD}{randomly-weighted and randomly-delayed}                   %129
\acrodef{SaS}[S$\alpha$S]{symmetric $\alpha$-stable}			%130

\acrodef{SA}{synthetic aperture}					%131
\acrodef{SBL}{sparse Bayesian learning}					%132
\acrodef{SDK}{software development kit}					%133
\acrodef{SDR}{software-defined radio}					%134
\acrodef{SIIM}{Society for Imaging Informatics in Medicine}		%135
\acrodef{SISO}{single-input, single-output}				%136
\acrodef{SINR}{signal-to-interference-plus-noise ratio}			%137
\acrodef{SMI}{sample matrix inversion}					%138
\acrodef{SNR}{signal-to-noise ratio}					%139
\acrodef{SOMP}{simultaneous orthogonal matching pursuit}		%140

\acrodef{SoS}{sum of sincs}						%141
\acrodef{SP}{subspace pursuit}						%142
\acrodef{spark}{sparse rank}						%143
\acrodef{SPECT}{single photon emission computed tomography}		%144
\acrodef{SPG}{spectral projected gradient}				%145
\acrodef{SPGL1}[SPG$\ell_{1}$]{spectral projected gradient for $\ell_{1}$-minimization}	%146
\acrodef{SPLS}{split-projection least squares}				%147
\acrodef{SR}{sparse recovery}						%148
\acrodef{SRC}{\name{Sommerfeld} radiation condition}			%149
\acrodef{SSI}{supersonic shear imaging}					%150

\acrodef{SSIM}{structural similarity}					%151
\acrodef{STFT}{short-time \name{Fourier} transform}			%152
\acrodef{StOMP}{stagewise \ac{OMP}}					%153
\acrodef{StWGP}{stagewise weak \ac{CGP}}				%154
\acrodef{SVD}{singular value decomposition}				%155
\acrodef{TDMA}{time-division multiple access}				%156
\acrodef{TMSBL}[T-MSBL]{temporally correlated \ac{MMV}-based \ac{SBL}}  %157
\acrodef{TOF}{time-of-flight}						%158.s
\acrodefplural{TOF}{times-of-flight}					%158.p
\acrodef{TGC}{time gain compensation}					%159
\acrodef{THI}{tissue harmonic imaging}					%160

\acrodef{TPSF}{transform point spread function}				%161
\acrodef{TPU}{tensor processing unit}                                   %162
\acrodef{TR}{\name{Tikhonov} regularization}				%163
\acrodef{TV}{total variation}						%164
\acrodef{UCA}{ultrasound contrast agents}				%165
\acrodef{UDT}{ultrasound diffraction tomography}			%166
\acrodef{UI}{ultrasound imaging}					%167
\acrodef{ULM}{ultrasound localization microscopy}			%168
\acrodef{UoS}{union of subspaces}					%169
\acrodef{USA}{United States of America}					%170

\acrodef{UWB}{ultra-wideband}						%171
\acrodef{voxel}{volume element}						%172
\acrodef{Xampling}{CS-Sampling}						%173

\acresetall

\begin{document}

%%%%%%%%%%%%%%%%%%%%%%%%%%%%%%%%%%%%%%%%%%%%%%%%%%%%%%%%%%%%%%%%%%%%%%%%%%%%%%%%%%%%%%%%%%%%%%%%%%%%%%%%%%%%%%%%
% results
%%%%%%%%%%%%%%%%%%%%%%%%%%%%%%%%%%%%%%%%%%%%%%%%%%%%%%%%%%%%%%%%%%%%%%%%%%%%%%%%%%%%%%%%%%%%%%%%%%%%%%%%%%%%%%%%
% Laplace-domain beamforming vs FBP
% median lateral FWHM: ( 533 - 499 ) * 100 / 533 = 6.378986867 %
\pgfmathsetmacro{\FWHMLateralMedianImprovement}{ 6.4 }
% median axial FWHM: ( 455 - 406 ) * 100 / 455 = 10.769230769 %
\pgfmathsetmacro{\FWHMAxialMedianImprovement}{ 10.8 }
% best wire (index: 2): ( 468.63 - 412.1979 ) * 100 / 468.63 = 12.041930734 %
\pgfmathsetmacro{\FWHMLateralImprovementUB}{ 12 }
% best wire (index: 3): ( 464.82 - 350.2521 ) * 100 / 464.82 = 24.647799148 %
\pgfmathsetmacro{\FWHMAxialImprovementUB}{ 25 }
% large region: 73 - 64 = 9
\pgfmathsetmacro{\gCNRImprovementUB}{ 9 }

%%%%%%%%%%%%%%%%%%%%%%%%%%%%%%%%%%%%%%%%%%%%%%%%%%%%%%%%%%%%%%%%%%%%%%%%%%%%%%%%%%%%%%%%%%%%%%%%%%%%%%%%%%%%%%%%
% copyright notice
%%%%%%%%%%%%%%%%%%%%%%%%%%%%%%%%%%%%%%%%%%%%%%%%%%%%%%%%%%%%%%%%%%%%%%%%%%%%%%%%%%%%%%%%%%%%%%%%%%%%%%%%%%%%%%%%
\begin{titlepage}
\thispagestyle{empty}%
\noindent
{\huge
Laplace-Domain Beamforming for\\[0.8ex]
Ultrafast Plane-Wave Imaging
}
\par
\vspace{36pt}
\noindent
{\large Martin F. Schiffner\par\vspace{12pt}
\noindent \href{http://www.mt.rub.de}{Chair of Medical Engineering}, Ruhr University Bochum, 44780 Bochum, Germany}
\vspace{36pt}
\par
\noindent
% see Section 8.1.9 of the PSPB Operations Manual:
% 8.1.9 Electronic Information Dissemination / B. COPYRIGHT NOTICE
{\bf Copyright notice:}\par\vspace{12pt}
\noindent
\copyright~2026~IEEE.
Personal use of
this material is
permitted.
Permission from
IEEE must be obtained for
all other uses, in
any current or
future media, including reprinting/republishing
this material for
advertising or
promotional purposes, creating
new collective works, for
resale or
redistribution to
servers or
lists, or
reuse of
any copyrighted component of
this work in
other works.
\par
\vspace{12pt}
\noindent
To be published in the proceedings of
the 2026 IEEE International Ultrasonics Symposium (IUS), Raleigh, NC, USA, Oct. 2026.
% {\bf Full citation:}\par\vspace{12pt}
% \noindent
% \par
% \noindent
% DOI: \href{http://dx.doi.org/???}{???}
% \par
% \vspace{12pt}
% \noindent
% \href{http://ieeexplore.ieee.org/document/???/}{Click here for IEEE Xplore}
\clearpage
\end{titlepage}

%%%%%%%%%%%%%%%%%%%%%%%%%%%%%%%%%%%%%%%%%%%%%%%%%%%%%%%%%%%%%%%%%%%%%%%%%%%%%%%%%%%%%%%%%%%%%%%%%%%%%%%%%%%%%%%%
% title, authors, and thanks
%%%%%%%%%%%%%%%%%%%%%%%%%%%%%%%%%%%%%%%%%%%%%%%%%%%%%%%%%%%%%%%%%%%%%%%%%%%%%%%%%%%%%%%%%%%%%%%%%%%%%%%%%%%%%%%%
% a) title Inversion
\title{Laplace-Domain Beamforming for Ultrafast Plane-Wave Imaging}

% b) list of authors
\author{%
  \IEEEauthorblockN{Martin F. Schiffner}%
  \IEEEauthorblockA{%
    \textit{Chair of Medical Engineering}\\
    \textit{Ruhr University Bochum}\\
    Bochum, Germany\\
%     martin.schiffner@rub.de%
    0000-0002-4896-2757\,\orcidlink{0000-0002-4896-2757}
  }
}

% c) create title
\maketitle

%%%%%%%%%%%%%%%%%%%%%%%%%%%%%%%%%%%%%%%%%%%%%%%%%%%%%%%%%%%%%%%%%%%%%%%%%%%%%%%%%%%%%%%%%%%%%%%%%%%%%%%%%%%%%%%%
% 0.) abstract
%%%%%%%%%%%%%%%%%%%%%%%%%%%%%%%%%%%%%%%%%%%%%%%%%%%%%%%%%%%%%%%%%%%%%%%%%%%%%%%%%%%%%%%%%%%%%%%%%%%%%%%%%%%%%%%%
\begin{abstract}
  %---------------------------------------------------------------------------------------------------------------
% 1.) abstract
%---------------------------------------------------------------------------------------------------------------
% acronyms:
% FBP: 2x
% FDT: 2x
% PW: 1x
% LDT: 1x
% a)
Fourier-domain beamforming methods, such as
% 1.) filtered backpropagation (FBP)
\ac{FBP}, are
% 2.) beneficial
beneficial in
% 3.) ultrafast plane-wave imaging
ultrafast plane-wave imaging due to
% 4.) computational efficiency
their computational efficiency and
% 5.) high image quality
high image quality.
% b)
These methods use
% 1.) Fourier diffraction theorem (FDT)
the \ac{FDT} for
% 2.) steered plane waves (PWs)
steered \aclp{PW} to exactly invert
% 3.) linear scattering models
linear scattering models.
% c)
However,
% 1.) linear scattering models
these models rely on
% 2.) simplifying assumptions
simplifying assumptions that degrade
% 3.) image quality
image quality.
% d)
The use of
% 1.) linear transducer arrays
linear transducer arrays, in particular, requires:
% 2.) two-dimensional space
(i)
a two-dimensional space, and
% 3.) lossless tissues
(ii)
lossless tissues.
% e)
Herein,
% 1.) Laplace-domain beamforming
Laplace-domain beamforming is
% 2.) proposed
proposed.
% f)
This method retains
% 1.) computational efficiency
the computational efficiency of
% 2.) Fourier-domain methods
the Fourier-domain methods while
% 3.) improving
improving
% 4.) image quality
image quality for
% 5.) linear arrays
linear arrays.
% g)
Specifically,
% 1.) method
the method uses
% 2.) generalization
a generalization of
% 3.) Fourier diffraction theorem (FDT)
the \ac{FDT}, here called
% 4.) Laplace diffraction theorem (LDT)
\acl{LDT}, to account for
% 5.) three spatial dimensions
three spatial dimensions,
% 6.) finite array element heights
finite array element heights, and
% 7.) lossy tissues
lossy tissues.
% h)
A phantom experiment showed that
% 1.) proposed method
the proposed method improves
% 3.) image uniformity
image uniformity,
% 4.) spatial resolution
spatial resolution, and
% 5.) contrast
contrast compared to
% 6.) filtered backpropagation (FBP)
\ac{FBP}.
% i)
Using
% 1.) three steering angles
three steering angles,
% 2.) lateral widths
the lateral and
% 3.) axial widths
axial \qty{-6}{\decibel}-widths of
% 4.) wires
wires reduced by
% 5.) up to \FWHMLateralImprovementUB %
up to \qty{\FWHMLateralImprovementUB}{\percent} and
% 6.) up to \FWHMAxialImprovementUB %
\qty{\FWHMAxialImprovementUB}{\percent},
respectively, while
% j)
the contrast of
% 1.) anechoic regions
anechoic regions improved by
% 2.) up to \gCNRImprovementUB %
up to \qty{\gCNRImprovementUB}{\percent}.

%   *CRITICAL: Do Not Use Symbols, Special Characters, Footnotes, or Math in Paper Title or Abstract.
\end{abstract}
\acresetall


%%%%%%%%%%%%%%%%%%%%%%%%%%%%%%%%%%%%%%%%%%%%%%%%%%%%%%%%%%%%%%%%%%%%%%%%%%%%%%%%%%%%%%%%%%%%%%%%%%%%%%%%%%%%%%%%
% 1.) introduction
%%%%%%%%%%%%%%%%%%%%%%%%%%%%%%%%%%%%%%%%%%%%%%%%%%%%%%%%%%%%%%%%%%%%%%%%%%%%%%%%%%%%%%%%%%%%%%%%%%%%%%%%%%%%%%%%
\section{Introduction}
\label{sec:introduction}
%---------------------------------------------------------------------------------------------------------------
% 1.) Fourier domain methods
%---------------------------------------------------------------------------------------------------------------
% a)
\name{Fourier}-domain beamforming methods, such as
\ac{FBP}
\cite{proc:SchiffnerAI2012},
% \cite[Sect. 8.9.2]{book:Devaney2012},
% \cite{article:DevaneyUltImag1982},
% 2.)
% article:LuITUFFC2006: High Frame Rate Imaging System for Limited Diffraction Array Beam Imaging with Square-Wave Aperture Weightings [Oct.]
% -
% article:ChengITUFFC2006: Extended high-frame rate imaging method with limited-diffraction beams [ May ]
% Abstract
% - In this paper, the theory is extended to include explicitly VARIOUS TRANSMISSION SCHEMES such as
%   MULTIPLE LIMITED-DIFFRACTION ARRAY BEAMS and
%   STEERED PLANE WAVES. (p. 880)
% - A relationship between the limited-diffraction array beam weighting of received echo signals and
%   a 2-D Fourier transform of the same signals over a transducer aperture is established. (p. 880)
% article:LuITUFFC1997: 2D and 3D High Frame Rate Imaging with Limited Diffraction Beams
% -
limited-diffraction beam imaging
\cite{article:ChengITUFFC2006}, and
% 3.) fast nonuniform inversion
% article:KruizingaITUFFC2012: Plane-wave ultrasound beamforming using a nonuniform fast Fourier transform
fast nonuniform inversion
\cite{article:KruizingaITUFFC2012}, are
% 4.) beneficial
beneficial in
% 5.) ultrafast plane-wave imaging
ultrafast plane-wave imaging due to
% 6.) computational efficiency
their computational efficiency and
% 7.) high image quality
high image quality.
% b)
These methods, unlike
% 1.) standard delay-and-sum approach
% article:SchiffnerUlt2024: Comment on "So you think you can DAS? A viewpoint on delay-and-sum beamforming" [Ultrasonics 111 (2021) 106309]
% article:PerrotUlt2021: So you think you can DAS? A viewpoint on delay-and-sum beamforming
the standard delay-and-sum approach, exploit
% 2.) Fourier diffraction theorem (FDT)
% book:Devaney2012: Mathematical Foundations of Imaging, Tomography and Waveﬁeld Inversion [Jul.]
% Chapter 8: Classical inverse scattering and diffraction tomography / Sect. 8.9 Diffraction tomography in two space dimensions
% Sect. 8.9.1 The generalized projection-slice theorem
% Theorem 8.4 (the generalized projection-slice theorem)
% (8.67)
% book:Natterer2004: (cf. e.g. [21, Thm. 8.4], [22, Thm. 3.1], [23, Sect. 6.3])
the \ac{FDT} for
% 3.) steered plane waves (PWs)
steered \acp{PW}
\cite[Eq. (8.67)]{book:Devaney2012}.
% c)
% a.k.a. generalized projection-slice theorem
This theorem enables
% 1.) exact inversion
exact inversion of
% 2.) linear scattering models
linear scattering models, expressing
% 3.) heterogeneous tissue properties
heterogeneous tissue properties directly as
% 4.) functions
functions of
% 5.) echo signals
the echo signals.
% d)
The inversion occurs in
% 1.) k-space
$k$-space
(i.e.,
  % 2.) domain
  the domain of
  % 3.) temporal frequencies
  temporal and
  % 4.) spatial frequencies
  spatial frequencies%
) and accounts for both
% 5.) phase information
phase and
% 6.) amplitude information
amplitude information.
% e)
The underlying scattering models, however, rely on
% 1.) simplifying assumptions | impair
simplifying assumptions that degrade
% 2.) image quality
image quality.
% f)
When using
% 1.) linear transducer arrays
linear transducer arrays,
% 2.) simplifying assumptions
these assumptions include:
% 3.) two-dimensional space
(i)
a two-dimensional space, and
% 4.) lossless tissues
(ii)
lossless tissues.
% g)
Both assumptions distort
% 1.) model-predicted amplitudes
the model-predicted amplitudes because
% 2.) attenuation
attenuation is
% 3.) underestimated
underestimated.

%---------------------------------------------------------------------------------------------------------------
% 2.) attenuation
%---------------------------------------------------------------------------------------------------------------
% a)
Attenuation limits
% 1.) imaging depth
imaging depth and degrades
% 2.) spatial resolution
spatial resolution as
% 3.) depth increases
depth increases.
% b)
% article:WellsUMB1975: Absorption and dispersion of ultrasound in biological tissue
% ATTENUATION AND ABSORPTION
% - When a wave travels through a medium,
%   its intensity is reduced as a function of distance. (p. 369)
% - Several factors contribute to this attenuation: they include
%   [1.)] BEAM DIVERGENCE,
%   [2.)] scattering,
%   [3.)] wave mode conversion and
%   [4.)] ABSORPTION, which is the process of conversion of vibrational energy into heat. (p. 369)
% Two primary mechanisms contributing to this loss are
Two contributing mechanisms are
% 1.) diffraction
diffraction and
% 2.) absorption
absorption.
% c)
% book:Devaney2012: Mathematical Foundations of Imaging, Tomography and Wavefield Inversion
% Chapter 2: Radiation and boundary-value problems in the frequency domain
% Sect. 2.2 Green functions / Sect. 2.2.1 Green functions in two space dimensions
% - The 2D Green functions are important since they reduce
%   the SIZE AND COMPLEXITY OF SIMULATION STUDIES in inverse scattering and diffraction tomography, and
%   will be employed extensively later in the book in such studies. (p. 53)
% Chapter 2: Radiation and boundary-value problems in the frequency domain
% Sect. 2.4 Green-function solution of the radiation problem / Sect. 2.4.1 Solution of the radiation problem in two space dimensions
% - We will make use of the 2D solution to the radiation problem in later chapters in
%   computer simulations and examples since it requires
%   MUCH LESS CPU TIME and IS EASIER TO IMPLEMENT than the 3D solution but retains
%   all of the salient features of the 3D case. (p. 58)
% Chapter 3: Eigenfunction expansions of solutions to the Helmholtz equation / Sect. 3.6 Two-dimensional wavefield
% - The 2D case is of importance for TESTING AND EVALUATING
%   WAVEFIELD PROPAGATION AND INVERSION ALGORITHMS in
%   COMPUTER SIMULATION STUDIES since it reduces
%   the computational burden to a minimum while still retaining
%   all of the flavor of the 3D case. (p. 113)
% - In addition, certain 3D problems reduce to a 2D formulation such as
%   scattering of plane waves off cylindrical structures. (p. 113)
A two-dimensional model underestimates
% 1.) diffraction-induced decay
the diffraction-induced decay of
% 2.) echo signals
the echo signals because
% 3.) elevational variations
elevational variations are
% 4.) absent
absent.
% d)
This absence implies that both
% 1.) array elements
the array elements and
% 2.) spatial fluctuations
spatial fluctuations in
% 3.) tissue properties
the tissue properties are of
% 4.) infinite height
infinite height.
% e)
The echoes, hence, are
% 1.) superpositions
superpositions of
% 2.) cylindrical waves
cylindrical waves whose
% 3.) amplitudes
amplitudes decay proportionally to
% 4.) 1 / \sqrt{ R }
$1 / \sqrt{ R }$ rather than
% 5.) 1 / R
$1 / R$, where
% 6.) R
$R$ denotes
% 7.) propagation distance
the propagation distance.
% f)
A lossless model, furthermore, ignores
% 1.) absorption
absorption, which exhibits
% 2.) power-law frequency dependence
a power-law frequency dependence and causes
% 3.) dispersion
dispersion.
% g)
% book:Cobbold2006, Chapter 3: Field Profile Analysis / Sect. 3.10 Effects of Attenuation
% Sect. 3.10.1 Kramers-Kronig Relationships
% - For acoustic waves,
%   the K-K equations can be expressed in terms of the components of
%   a COMPLEX PROPAGATION CONSTANT that can be defined by
%   [ \munderbar{k}( \omega ) = \beta( \omega ) - j \alpha( \omega ) = \frac{ \omega }{ c_{0}( \omega ) } - j \alpha( \omega ), ] (3.95)
%   in which
%   c_{0}( \omega ) denotes the phase speed and
%   \alpha( \omega ) is the attenuation coefficient. (p. 206)
Both effects are
% 1.) encapsulated
encapsulated by
% 2.) complex-valued wavenumber
a complex wavenumber
\cite[Eq. (3.95)]{book:Cobbold2006}
\begin{equation}
 %--------------------------------------------------------------------------------------------------------------
 % complex-valued wavenumber
 %--------------------------------------------------------------------------------------------------------------
  \munderbar{k}( f )
  =
  \beta( f )
  -
  j
  \alpha( f ),
 \label{eqn:intro_wavenumber}
\end{equation}
where
% 3.) f
$f$ denotes
% 4.) frequency
the frequency,
% 5.) j
$j$ is
% 6.) imaginary unit
the imaginary unit,
% 7.) \beta( f )
$\beta( f ) = 2 \pi f / c( f )$ is
% 8.) phase constant
the phase constant associated with
% 9.) phase velocity
the phase velocity
$c( f )$, and
% 10.) \alpha( f )
$\alpha( f )$ is
% 11.) spatial amplitude absorption coefficient
the absorption coefficient.

%%%%%%%%%%%%%%%%%%%%%%%%%%%%%%%%%%%%%%%%%%%%%%%%%%%%%%%%%%%%%%%%%%%%%%%%%%%%%%%%%%%%%%%%%%%%%%%%%%%%%%%%%%%%%%%%
% graphic: measurement setup
%%%%%%%%%%%%%%%%%%%%%%%%%%%%%%%%%%%%%%%%%%%%%%%%%%%%%%%%%%%%%%%%%%%%%%%%%%%%%%%%%%%%%%%%%%%%%%%%%%%%%%%%%%%%%%%%
%
\begin{figure}[t!]
 \centering%
  \input{theory/figures/latex/theory_measurement_setup.tex}
 \vspace*{0mm}
 \caption{}
 \label{fig:V}
\end{figure}
C
{% a)
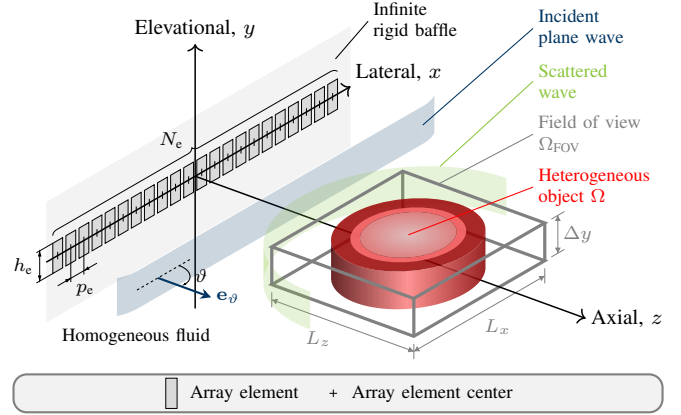
 Model of
 % 1.) pulse-echo measurement setup
 the measurement setup.
 % b)
 The \acs{UI} system uses
 % 1.) linear transducer array
 a linear transducer array to transmit
 % 2.) steered plane waves (PWs)
 steered \acp{PW} and record
 % 3.) received echoes
 the received echoes.
 % c)
 The array elements are
 % 1.) embedded
 embedded in
 % 2.) infinite rigid baffle
 an infinite rigid baffle on
 % 3.) plane z = 0
 the plane
 $z = 0$.
 % d)
 The transmitted \acp{PW} propagate through
 % 1.) lossy fluid
 a lossy fluid that occupies
 % 2.) positive half-space z > 0 | right
 the positive half-space
 $z > 0$ and is
 % 3.) homogeneous
 homogeneous everywhere except within
 % 4.) bounded set
 the bounded set
 $\Omega \subset \{ \vect{r} \in \R^{3}: z > 0 \}$.
 % e)
 This set describes
 % 1.) heterogeneous object
 a heterogeneous object whose
 % 2.) compressibility
 compressibility
 $\kappa_{1}( \vect{r} )$ fluctuates
 % 3.) spatially
 spatially.
 % f)
 These fluctuations, upon
 % 1.) insonification
 insonification by
 % 2.) incident plane wave (PW)
 an incident \ac{PW}, induce
 % 3.) scattered waves
 scattered waves that propagate back to
 % 4.) planar transducer array
 the array and mechanically excite
 % 5.) array elements
 its elements.
}%
{theory_measurement_setup}

%---------------------------------------------------------------------------------------------------------------
% 3.) contents
%---------------------------------------------------------------------------------------------------------------
% a)
Herein,
% 1.) novel beamforming method
a novel beamforming method termed
% 2.) Laplace-domain beamforming
\term{\name{Laplace}-domain beamforming} is
% 3.) proposed
proposed.
% b)
This method retains
% 1.) computational efficiency
the computational efficiency of
% 2.) Fourier-domain methods
\name{Fourier}-domain methods while improving
% 3.) image quality
image quality for
% 4.) linear arrays
linear arrays.
% c)
Specifically,
% 1.) proposed method
the method relaxes
% 2.) above constraints
the above constraints of
% 3.) Fourier diffraction theorem (FDT)
the \ac{FDT} by accounting for
% 4.) three-dimensional spatial mechanics,
three dimensions,
% 5.) finite array element heights
finite array element heights, and
% 6.) lossy tissues
lossy tissues.
% d)
These modifications yield
% 1.) generalization | extension | generalized extension
a generalization of
% 2.) Fourier diffraction theorem (FDT)
the \ac{FDT}, here designated as
% 3.) Laplace diffraction theorem (LDT)
the \term{\acl{LDT}} (\acs{LDT})\acused{LDT}.
% e)
This theorem relates
% 1.) transverse spectrum
the transverse spectrum of
% 2.) scattered acoustic pressure field
the scattered acoustic pressure field
% 3.) induced
induced by
% 4.) steered plane wave (PW)
an incident \ac{PW} to
% 5.) spatial Laplace transform
the spatial \name{Laplace} transform of
% 6.) tissue properties
the tissue properties at
% 7.) specific complex spatial frequencies
specific complex frequencies.
% f)
This relationship, modeling
% 1.) tissue properties
the tissue properties within
% 2.) field of view (FOV)
the \ac{FOV} as
% 3.) Fourier sum
a \name{Fourier} sum, yields
% 4.) compressible linear equation system
a linear equation system that can be
% 5.) solved efficiently
solved efficiently using
% 6.) standard regularization methods
standard regularization methods to form
% 7.) images
images.

%%%%%%%%%%%%%%%%%%%%%%%%%%%%%%%%%%%%%%%%%%%%%%%%%%%%%%%%%%%%%%%%%%%%%%%%%%%%%%%%%%%%%%%%%%%%%%%%%%%%%%%%%%%%%%%%
% 2.) theory
%%%%%%%%%%%%%%%%%%%%%%%%%%%%%%%%%%%%%%%%%%%%%%%%%%%%%%%%%%%%%%%%%%%%%%%%%%%%%%%%%%%%%%%%%%%%%%%%%%%%%%%%%%%%%%%%
\section{Theory}
% \label{sec:theory}
%---------------------------------------------------------------------------------------------------------------
% 1.) theory
%---------------------------------------------------------------------------------------------------------------
% a)
The following derivations, as shown in
% 1.) Figure 1
\cref{fig:theory_measurement_setup}, apply to
% 2.) plane-wave imaging
plane-wave imaging with
% 3.) linear array
a linear array
(%
  % 3.a) number of elements
  number of elements: $N_{\text{e}}$,
  % 3.b) pitch
  pitch: $p_{\text{e}}$,
  % 3.c) element height
  element height: $h_{\text{e}}$%
) in
% 4.) three-dimensional space
three-dimensional space.
% b)
Any position is
% 1.) specified
specified by
% 2.) vector
the vector
$\vect{r} = \trans{ [ x, y, z ] }$, where
% 3.) \vect{r}_{\rho}
$\vect{r}_{\rho} = \trans{ [ x, y ] }$ denotes
% 4.) transverse position
the transverse position and
% 5.) z
$z$ is
% 6.) axial position
the axial position.
% c)
The aperture of
% 1.)
the linear array is
% 2.)
embedded in
% 3.)
an infinite rigid baffle on
% 4.)
the plane
$z = 0$.
% d)
The positive half-space
$z > 0$ is
% 1.) lossy fluid
a lossy fluid whose
% 2.) compressibility
compressibility
$\kappa_{0}$ is
% 3.) homogeneous
homogeneous except within
% 4.) bounded volume
the bounded set
$\Omega \subset \{ \vect{r} \in \R^{3}: z > 0 \}$.
% e)
This set contains
% 1.) heterogeneous object
a heterogeneous object with
% 2.) compressibility
compressibility
$\kappa_{1}( \vect{r} )$.
% f)
The object function
$\gamma( \vect{r} )$ thus reads
\begin{equation}
 %--------------------------------------------------------------------------------------------------------------
 % relative spatial fluctuations in the unperturbed compressibility [ object function ]
 %--------------------------------------------------------------------------------------------------------------
  \gamma( \vect{r} )
  =
  \begin{cases}
   %------------------------------------------------------------------------------------------------------------
   % a) heterogeneous object
   %------------------------------------------------------------------------------------------------------------
    1
    -
    \kappa_{1}( \vect{r} ) / \kappa_{0}
  & \text{for } \vect{r} \in \Omega,\\
   %------------------------------------------------------------------------------------------------------------
   % b) homogeneous fluid
   %------------------------------------------------------------------------------------------------------------
    0
  & \text{for } \vect{r} \notin \Omega.
  \end{cases}
 \label{eqn:theory_rel_fluctuations}
\end{equation}
% g)
% article:WilcoxITUFFC2018: Quantification of the Effect of Array Element Pitch on Imaging Performance
% II. THEORETICAL DEVELOPMENT / A. General Framework
% - It will be assumed here that f_{ij}( t ) is analytic
%   (i.e., it comprises both
%   [1.)] the real physically recorded data and
%   [2.)] an imaginary component in quadrature as obtained through the Hilbert transform). (p. 601)
% book:Szabo2013, Appendix A: The Fourier Transform / Sect. A.2 The Fourier Transform
% Sect. A.2.7 Calculating an Inverse Fourier Transform and a Hilbert Transform with an FFT
% - The overall time signal that consists of
%   [1.)] the real and
%   [2.)] imaginary quadrature signals is called the “analytic signal.” (p. 781)
All acoustic fields are
% 1.)
% assumed to be
% 2.) analytic \cite[Sect. A.2.7]{book:Szabo2013} 
% analytic and
% 3.) bandpass bandlimited
% article:HerleyITIT1999: Minimum Rate Sampling and Reconstruction of Signals with Arbitrary Frequency Support
% I. INTRODUCTION
% - The idea that a signal which has
%   ENERGY IN ONLY A LIMITED RANGE OF FREQUENCIES may be represented by its samples is
%   of fundamental importance in digital communications. (p. 1555)
% - In the version given by Shannon it is stated that
%   a real signal that has ENERGY ONLY IN THE RANGE OF FREQUENCIES ( - B, B ) can be reconstructed without loss from
%   uniform samples at rate 2B. (p. 1555)
bandpass bandlimited
(i.e.,
  there are
  % 3.a) lower frequency bound
  lower and
  % 3.b) upper frequency bound
  upper bounds
  % 4.a) f_{\text{lb}} > 0
  $f_{\text{lb}} > 0$ and
  % 4.b) f_{\text{ub}} > f_{\text{lb}}
  $f_{\text{ub}} > f_{\text{lb}}$, respectively, such that
  % 5.) fields
  the fields have
  % 6.) energy
  energy only in
  % 7.) range
  the range of
  % 8.) frequencies
  frequencies
  $f_{\text{lb}} \leq f \leq f_{\text{ub}}$%
).
% h)
However, unless
% 1.) indicated otherwise
indicated otherwise,
% 2.) only a single frequency
only a single frequency is %in
% 3.) single pulse-echo measurement
% a single pulse-echo measurement is
% 4.) considered
considered, and
% 5.) argument
the argument
$f$ is
% 6.) omitted
omitted to simplify
% 7.) presentation
the notation.
% j)
The incident \ac{PW} is
% 1.)
specified first.
% k)
Next,
% 1.) proposed Laplace diffraction theorem (LDT)
the proposed \ac{LDT} is
% 2.) established
established, and
% 3.) element height
the element height
$h_{\text{e}}$ is
% 4.) incorporated
incorporated.
% l)
The resulting equations are
% 1.) discretized
discretized and, lastly, exploited for
% 2.) image formation
image formation.

%%%%%%%%%%%%%%%%%%%%%%%%%%%%%%%%%%%%%%%%%%%%%%%%%%%%%%%%%%%%%%%%%%%%%%%%%%%%%%%%%%%%%%%%%%%%%%%%%%%%%%%%%%%%%%%%
% 1.) incident wave
%%%%%%%%%%%%%%%%%%%%%%%%%%%%%%%%%%%%%%%%%%%%%%%%%%%%%%%%%%%%%%%%%%%%%%%%%%%%%%%%%%%%%%%%%%%%%%%%%%%%%%%%%%%%%%%%
\subsection{Incident Wave}
% \label{subsec:theory_incident_wave}
%---------------------------------------------------------------------------------------------------------------
% 1.) incident field
%---------------------------------------------------------------------------------------------------------------
% a)
The incident wave is
% 1.) steered plane wave (PW)
a steered \ac{PW} with
% 2.) steering angle
the steering angle
$\vartheta \in ( 0; \pi )$.
% b)
This wave has
% 1.) acoustic pressure field
the acoustic pressure field
\begin{subequations}
\label{eqn:theory_p_in} 
\begin{equation}
 %--------------------------------------------------------------------------------------------------------------
 % incident acoustic pressure field [ steered PW ]
 %--------------------------------------------------------------------------------------------------------------
  p^{(\text{in})}( \vect{r} )
  =
  A^{(\text{in})}
  e^{ -j \munderbar{k} \inprod{ \uvect{\vartheta} }{ \vect{r} - \vect{r}_{\text{ref}} } },
 \label{eqn:theory_p_in_expression}
\end{equation}
where
% 2.) A^{(\text{in})}
$A^{(\text{in})}$ is
% 3.) complex amplitude
the complex amplitude,
% 4.) \munderbar{k}
$\munderbar{k}$ is
% 5.) complex-valued wavenumber
the complex wavenumber
\eqref{eqn:intro_wavenumber}, and
% 6.)
$\tinprod{ \uvect{\vartheta} }{ \vect{r} - \vect{r}_{\text{ref}} }$ is
% 7.) scalar product
the scalar product of
% 8.) propagation direction
the propagation direction
$\uvect{\vartheta} = \trans{ [ \cos( \vartheta ), 0, \sin( \vartheta ) ] }$ and
% 9.) difference
the difference of
% 10.) position
the position
$\vect{r}$ and
% 11.) reference positions for the time delays and their components
the reference position
$\vect{r}_{\text{ref}} = \trans{ [ x_{\text{ref}}, 0, 0 ] }$ with
% 12.) lateral component
the lateral component
\begin{equation}
 %--------------------------------------------------------------------------------------------------------------
 % components of the reference positions for the time delays
 %--------------------------------------------------------------------------------------------------------------
  x_{\text{ref}}
  =
  \begin{cases}
   %------------------------------------------------------------------------------------------------------------
   % a) leftmost element
   %------------------------------------------------------------------------------------------------------------
    - ( N_{\text{e}} - 1 ) p_{\text{e}} / 2 &
    \text{for } \vartheta \leq \pi / 2,\\
   %------------------------------------------------------------------------------------------------------------
   % b) rightmost element
   %------------------------------------------------------------------------------------------------------------
      ( N_{\text{e}} - 1 ) p_{\text{e}} / 2 &
    \text{for } \vartheta > \pi / 2.
  \end{cases}
 \label{eqn:app_fdt_pw_p_in_r_ref}
\end{equation}
\end{subequations}
% c) \TODO{besser!}
This position ensures
% 1.) causality
causality of
% 2.) wave emission
the wave emission.

%%%%%%%%%%%%%%%%%%%%%%%%%%%%%%%%%%%%%%%%%%%%%%%%%%%%%%%%%%%%%%%%%%%%%%%%%%%%%%%%%%%%%%%%%%%%%%%%%%%%%%%%%%%%%%%%
% 2.) Laplace diffraction theorem for steered plane waves
%%%%%%%%%%%%%%%%%%%%%%%%%%%%%%%%%%%%%%%%%%%%%%%%%%%%%%%%%%%%%%%%%%%%%%%%%%%%%%%%%%%%%%%%%%%%%%%%%%%%%%%%%%%%%%%%
\subsection{Laplace Diffraction Theorem for Steered Plane Waves}
% \label{subsec:theory_f_number_transmit}
%%%%%%%%%%%%%%%%%%%%%%%%%%%%%%%%%%%%%%%%%%%%%%%%%%%%%%%%%%%%%%%%%%%%%%%%%%%%%%%%%%%%%%%%%%%%%%%%%%%%%%%%%%%%%%%%
% graphic: passbands predicted by Laplace diffraction theorem (LDT)
%%%%%%%%%%%%%%%%%%%%%%%%%%%%%%%%%%%%%%%%%%%%%%%%%%%%%%%%%%%%%%%%%%%%%%%%%%%%%%%%%%%%%%%%%%%%%%%%%%%%%%%%%%%%%%%%
%
\begin{figure}[t!]
 \centering%
  \input{theory/figures/latex/theory_passbands_pw.tex}
 \vspace*{-3mm}
 \caption{}
 \label{fig:V}
\end{figure}
C
{% a)
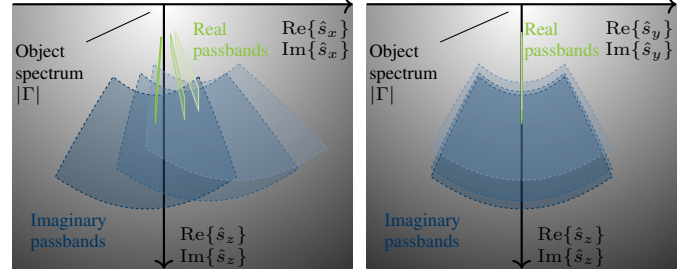
 Illustration of
 % 1.) proposed Laplace diffraction theorem (LDT)
 the proposed \ac{LDT}.
 % b)
 This theorem shows that
 % 1) pulse-echo measurement
 each pulse-echo measurement acts as
 % 2.) bandpass filter
 a bandpass filter for
 % 3.) relative spatial fluctuations in the unperturbed compressibility
 the object function
 \eqref{eqn:theory_rel_fluctuations}.
 % c)
 For
 \subref{fig:theory_passbands_pw_mono}
 % 1.) monofrequent incident plane wave (PW)
 the incident \ac{PW} with
 % 2.) incident acoustic pressure field [ steered PW ]
 the acoustic pressure field
 \eqref{eqn:theory_p_in},
 % 3.) complex angular spatial frequencies
 the complex angular spatial frequencies
 \eqref{eqn:theory_ldt_spatial_frequencies} describe
 % 4.) two structures
 two structures.
 % d)
 The real part
 $\treal{ \vect{T}_{\vartheta}( \vect{k}_{\rho} ) }$ describes
 % 1.) line
 a line in
 % 2.) \treal{ \hat{s}_{x} } \treal{ \hat{s}_{z} }-plane
 the $\treal{ \hat{s}_{x} } \treal{ \hat{s}_{z} }$-plane
 (i.e., $\treal{ \hat{s}_{y} } = 0$) parallel to
 % 3.) \hat{s}_{z}-axis
 the $\treal{ \hat{s}_{z} }$-axis
 (green dashed).
 % e)
 The imaginary part
 $\timag{ \vect{T}_{\vartheta}( \vect{k}_{\rho} ) }$, in contrast, resembles
 % 1.) hemisphere
 a hemisphere with
 % 2.) center
 the center
 $\beta \uvect{\vartheta}$ and
 % 3.) radius
 the radius
 $\sigma$
 (blue dashed).
 % f)
 For
 % 1.) sequence
 \subref{fig:theory_passbands_pw_broadband_sequence}
 a sequence of
 % 2.) three steered plane waves (PWs)
 three steered \acp{PW} in
 % 3.) frequency range
 the frequency range
 $f_{\text{lb}} \leq f \leq f_{\text{ub}}$,
 % 4.) complex angular spatial frequencies
 the complex angular spatial frequencies
 \eqref{eqn:theory_ldt_spatial_frequencies} describe
 % 5.) illustrated structures
 the illustrated structures.
 % g)
 % proc:SchiffnerIUS2026: Laplace-Domain Beamforming for Ultrafast Plane-Wave Imaging
 % III. METHODS / A. Image Formation (subsec:methods_image_formation)
 % - The \name{Laplace}-domain methods used
 %   the complex time-causal wavenumber \cite[Eq. (3.98)]{book:Cobbold2006}.
 The complex wavenumber
 % 1.) complex-valued wavenumber
 \eqref{eqn:intro_wavenumber}, in
 % 2.) both cases
 both cases, is
 % 3.) computed
 computed using
 % 4.) time-causal model
 the time-causal model
 (see \cref{subsec:methods_image_formation}).
 % h)
 To enlarge
 % 1.) real part
 the real part
 $\treal{ \vect{T}_{\vartheta}( \vect{k}_{\rho} ) }$ relative to
 % 2.) imaginary part
 the imaginary part
 $\timag{ \vect{T}_{\vartheta}( \vect{k}_{\rho} ) }$,
 % 3.) absorption parameters
 the absorption parameters were
 % 4.) set
 set to yield
 $\alpha( f_{\text{ub}} ) / \beta( f_{\text{ub}} ) \approx 0.5$.
 % i)
 This quotient is usually
 % 1.) much smaller
 much smaller. % in
 % 2.) soft tissues
%  soft tissues.
}%
{theory_passbands}

%---------------------------------------------------------------------------------------------------------------
% 1.) Laplace diffraction theorem for steered plane waves
%---------------------------------------------------------------------------------------------------------------
% a)
The proposed \ac{LDT}, as illustrated in
% 1.) Figure 2
\cref{fig:theory_passbands}, shows that
% 2.) pulse-echo measurement
each pulse-echo measurement acts as
% 3.) bandpass filter
a bandpass filter for
% 4.) relative spatial fluctuations in the unperturbed compressibility
the object function
\eqref{eqn:theory_rel_fluctuations}.
% b)
% In fact, | indeed
The \name{Born} approximation of
% 1.) scattered acoustic pressure field
the scattered acoustic pressure field on
% 2.) aperture plane
the aperture plane
(i.e., $z = 0$) contains
% 3.) information
information about
% 4.) specific angular spatial frequencies
specific angular spatial frequencies.
% c)
These frequencies, denoted by
% 1.) denoted
$\hat{\vect{s}} = \trans{ [ \hat{s}_{x}, \hat{s}_{y}, \hat{s}_{z} ] }$, are
% 2.) complex-valued | are revealed by expanding | manifest by expanding | manifest through expanding | can be represented by expanding | can be expressed by expanding
complex and manifest by
% 3.) expanding
expanding
% 4.) Born approximation of the scattered acoustic pressure field
the field as
% 5.) superposition
a superposition of
% 6.) steered plane waves (PWs)
steered \acp{PW}.
% d)
Each such wave has
% 1.) angular spatial frequency vector
an angular spatial frequency vector
$\vect{k}( \vect{k}_{\rho} ) = \trans{ [ \trans{ \vect{k}_{\rho} }, \munderbar{k}_{z}( \vect{k}_{\rho} ) ] }$.
% e)
% 3.) transverse frequencies
% article:SchiffnerITUFFC2023: Frequency-Dependent F-Number Suppresses Grating Lobes and Improves the Lateral Resolution in Coherent Plane-Wave Compounding
% APPENDIX I
%  square of complex-valued wavenumber:
%
%    \munderbar{k}^{2}
%    =
%    ( \beta - j \alpha )^{2}
%    =
%    \beta^{2} - \alpha^{2} - j 2 \alpha \beta 
%    =
%    \beta^{2} [ 1 - ( \alpha / \beta )^{2} ] - j 2 \alpha \beta
The transverse frequencies
$\vect{k}_{\rho} = \trans{ [ k_{x}, k_{y} ] }$ are
% 1.) real
real and determine
% 2.) complex angular axial frequency
the complex axial frequency
$\munderbar{k}_{z}( \vect{k}_{\rho} )$ via
% 3.) complex square root
the square root
\cite[Appendix I]{article:SchiffnerITUFFC2023}
\begin{equation*}
 %--------------------------------------------------------------------------------------------------------------
 % normalized complex-valued square root
 %--------------------------------------------------------------------------------------------------------------
  \munderbar{k}_{z}( \vect{k}_{\rho} )
%   =
%   \sqrt{ \munderbar{k}^{2} - \norm{ \vect{k}_{\rho} }{2}^{2} }
%   =
%   \sqrt{ \beta^{2} [ 1 - ( \alpha / \beta )^{2} ] - j 2 \alpha \beta - \norm{ \vect{k}_{\rho} }{2}^{2} }
%   =
%   \sigma
%   \sqrt{ 1 - \frac{ \norm{ \vect{k}_{\rho} }{2}^{2} }{ \sigma^{2} } - j \frac{ 2 \alpha \beta }{ \sigma^{2} } }
  =
  \sigma
  \sqrt{ 1 - \frac{ \norm{ \vect{k}_{\rho} }{2}^{2} }{ \sigma^{2} } - j \frac{ 2 \alpha / \beta }{ 1 - ( \alpha / \beta )^{2} } }
\end{equation*}
with
% 4.) reference frequency for normalization
$\sigma = \beta \sqrt{ 1 - ( \alpha / \beta )^{2} }$ and
% 5.) imaginary part
$\timag{ \munderbar{k}_{z}( \vect{k}_{\rho} ) } \leq 0$.
% f)
% evanescent PWs:
%
% \bar{k}_{z}( \bar{\vect{k}}_{\rho} )
% =
% \sqrt{ 1 - \norm{ \bar{\vect{k}}_{\rho} }{2}^{2} - j 2 \alpha \beta / \sigma^{2} }
% =
% \sqrt{ 1 - \norm{ \bar{\vect{k}}_{\rho} }{2}^{2} }
% \sqrt{ 1 - j 2 \alpha \beta / ( \sigma^{2} ( 1 - \norm{ \bar{\vect{k}}_{\rho} }{2}^{2} ) ) }
% \approx
% \frac{
%   \alpha \beta
% }{
%   \sigma^{2}
%   \sqrt{ \norm{ \bar{\vect{k}}_{\rho} }{2}^{2} - 1 }
% }
% -
% j
% \sqrt{ \norm{ \bar{\vect{k}}_{\rho} }{2}^{2} - 1 }
%
% propagation
% e^{ -j \sigma \bar{k}_{z}( \bar{\vect{k}}_{\rho} ) z_{0} }
% =
% e^{ -j \alpha \beta / ( \sigma \sqrt{ \norm{ \bar{\vect{k}}_{\rho} }{2}^{2} - 1 } ) z_{0} }
% e^{ - \sigma \sqrt{ \norm{ \bar{\vect{k}}_{\rho} }{2}^{2} - 1 } z_{0} }
If
% 1.) transverse frequencies
the transverse frequencies
$\vect{k}_{\rho}$ are
% 2.) large | outside the open disk
large
(i.e.,
  % 3.)
  $\vect{k}_{\rho} \notin \opendisk{ \sigma }$, where
  % 4.) \opendisk{ \sigma }
  $\opendisk{ \sigma } = \{ \vect{k}_{\rho} \in \R^{2}: \tnorm{ \vect{k}_{\rho} }{2} < \sigma \}$ denotes
  % 5.) open disk
  the open disk of
  % 6.) radius
  radius
  $\sigma$%
),
% 7.) plane waves (PWs)
the \acp{PW} are
% 8.) evanescent
evanescent and may be
% 9.) neglected
neglected.
% g)
% homogeneous PWs / weakly inhomogeneous PWs:
%
% \bar{k}_{z}( \bar{\vect{k}}_{\rho} )
% =
% \sqrt{ 1 - \norm{ \bar{\vect{k}}_{\rho} }{2}^{2} - j 2 \alpha \beta / \sigma^{2} }
% =
% \sqrt{ 1 - \norm{ \bar{\vect{k}}_{\rho} }{2}^{2} }
% \sqrt{ 1 - j 2 \alpha \beta / ( \sigma^{2} ( 1 - \norm{ \bar{\vect{k}}_{\rho} }{2}^{2} ) ) }
% \approx
% \sqrt{ 1 - \norm{ \bar{\vect{k}}_{\rho} }{2}^{2} }
% -
% j
% \frac{
%   \alpha \beta
% }{
%   \sigma^{2}
%   \sqrt{ 1 - \norm{ \bar{\vect{k}}_{\rho} }{2}^{2} }
% }
%
% propagation:
% e^{ -j \sigma \bar{k}_{z}( \bar{\vect{k}}_{\rho} ) z_{0} }
% =
% e^{ -j \sigma \sqrt{ 1 - \norm{ \bar{\vect{k}}_{\rho} }{2}^{2} } z_{0} }
% e^{ - \alpha \beta / ( \sigma \sqrt{ 1 - \norm{ \bar{\vect{k}}_{\rho} }{2}^{2} } ) z_{0} }
%
% complex-valued wave vector:
% \vect{k}( \bar{\vect{k}}_{\rho} )
% =
% \sigma
% \trans{ \left[ \bar{\vect{k}}_{\rho}, - \bar{k}_{z}( \bar{\vect{k}}_{\rho} ) \right] }
% \approx
% \sigma
% (
%   \trans{ \left[ \bar{\vect{k}}_{\rho}, - \sqrt{ 1 - \norm{ \bar{\vect{k}}_{\rho} }{2}^{2} } \right] }
%   +
%   j
%   \trans{ \left[ \vect{ 0 }, \frac{ \alpha \beta }{ \sigma^{2} \sqrt{ 1 - \norm{ \bar{\vect{k}}_{\rho} }{2}^{2} } } \right] }
% )
Otherwise,
% 1.) plane waves (PWs)
the \acp{PW} are
% 2.) propagable
propagable and, considering
% 3.) incident plane wave (PW)
the incident \ac{PW} with
% 4.) incident acoustic pressure field [ steered PW ]
the acoustic pressure field
\eqref{eqn:theory_p_in}, carry
% 5.) information
information about
% 6.) complex angular spatial frequencies
%  frequencies:
%
%    \vect{T}_{\vartheta}( \vect{k}_{\rho} )
%    =
%    j
%    \bigl[
%      \munderbar{k}
%      \uvect{\vartheta}
%      +
%      \vect{k}( \vect{k}_{\rho} )
%    \bigr]
the complex angular spatial frequencies
\begin{equation}
 %--------------------------------------------------------------------------------------------------------------
 % complex-valued angular spatial frequencies [steered PWs]
 %--------------------------------------------------------------------------------------------------------------
  \hat{\vect{s}}
  =
  \vect{T}_{\vartheta}( \vect{k}_{\rho} )
  =
  \begin{bmatrix}
    T_{\vartheta, x}( k_{x} )\\
    T_{\vartheta, y}( k_{y} )\\
    T_{\vartheta, z}( \vect{k}_{\rho} )
  \end{bmatrix}
  =
  j
  \bigl[
    \munderbar{k}
    \uvect{\vartheta}
    +
    \vect{k}( \vect{k}_{\rho} )
  \bigr].
%   =
%   j
%   \begin{bmatrix}
%     \munderbar{k} \cos( \vartheta ) + k_{x}\\
%     k_{y}\\
%     \munderbar{k} \sin( \vartheta ) + \munderbar{k}_{z}( \vect{k}_{\rho} )
%   \end{bmatrix}.
%   =
%   \alpha
%   \uvect{\vartheta}
%   -
%   \imag{ \vect{k}( \vect{k}_{\rho} ) }
%   +
%   j
%   \bigl[
%     \beta
%     \uvect{\vartheta}
%     +
%     \real{ \vect{k}( \vect{k}_{\rho} ) }
%   \bigr]
%   =
%   \alpha
%   \uvect{\vartheta}
%   -
%   \begin{pmatrix}
%     \vect{0} \\
%     \imag{ \munderbar{k}_{z}( \vect{k}_{\rho} ) }
%   \end{pmatrix}
%   +
%   j
%   \left[
%     \beta
%     \uvect{\vartheta}
%     +
%     \begin{pmatrix}
%       \vect{k}_{\rho} \\
%       \real{ \munderbar{k}_{z}( \vect{k}_{\rho} ) }
%     \end{pmatrix}
%   \right]
 \label{eqn:theory_ldt_spatial_frequencies}
\end{equation}
The above field, in particular, has
% 1.) plane-wave expansion
the expansion
\begin{equation}
 %--------------------------------------------------------------------------------------------------------------
 % Born approximation of the scattered acoustic pressure field [ compressibility; Fourier domain; plane-wave decomposition; rigid baffle ]
 %--------------------------------------------------------------------------------------------------------------
  p^{(\text{sc})}( \vect{r}_{\rho} )
%   =
%   \frac{
%     j \munderbar{k}
%     p^{(\text{in})}( \vect{0} )
%   }{
%     ( 2 \pi )^{2}
%   }
%   \int_{ \opendisk{ \sigma } }
%     \frac{
%       \Gamma\left[ \vect{T}_{\vartheta}( \vect{k}_{\rho} ) \right]
%       e^{ j \dinprod{ \vect{k}_{\rho} }{ \vect{r}_{\rho} }{0} }
%     }{
%       Z_{0} / Z_{2}
%       +
%       \munderbar{k}_{z}( \vect{k}_{\rho} ) / \munderbar{k}
%     }
%   \text{d} \vect{k}_{\rho},
  =
  \frac{
    j \munderbar{k}^{2}
    p^{(\text{in})}( \vect{0} )
  }{
    ( 2 \pi )^{2}
  }
  \int_{ \opendisk{ \sigma } }
    \frac{
      \Gamma\left[ \vect{T}_{\vartheta}( \vect{k}_{\rho} ) \right]
      e^{ j \dinprod{ \vect{k}_{\rho} }{ \vect{r}_{\rho} }{0} }
    }{
      \munderbar{k}_{z}( \vect{k}_{\rho} )
    }
  \text{d} \vect{k}_{\rho},
 \label{eqn:theory_p_sc}
\end{equation}
where
the plane-wave decomposition of
% 2.) Green's function
the \name{Green}'s function
\cite[(4.4)]{book:Devaney2012} was used, and
% 3.) \Gamma
$\Gamma$ denotes
% 4.) spatial Laplace transform
% book:Cohen2007, Sect. 1.8: Multidimensional Laplace Transforms
% - We have restricted our exposition of LAPLACE TRANSFORMS to
%   functions of one variable but we could equally well EXTEND OUR DEFINITION to
%   a FUNCTION OF TWO OR MORE VARIABLES. (p. 18)
% - Thus if, for example, y is a function of the variables t1, t2, i.e. y = f( t1, t2 ) then we define
%   [...] (1.38), where
%   for convergence of the integrals, we have to impose restrictions on
%   f and s1, s2 which are similar to those in §1.1,
%   — see Ditkin and Prudnikov [67] for a fuller account of the criteria needed for convergence. (p. 18)
% article:LubbockPIEE1969: Multidimensional Laplace transforms for solution of nonlinear equations
% - The multidimensional transforms of y(t) and x(t) can be defined as
%   [...] (4).
%
%  Laplace transform:
%
%    \Gamma( \hat{\vect{s}} )
%    =
%    \int_{ \Omega }
%      \gamma( \vect{r} )
%      e^{ - \dinprodr{ \hat{\vect{s}} }{ \vect{r} }{0} }
%    \text{d} \vect{r}
the spatial \name{Laplace} transform of
% 5.) relative spatial fluctuations in the unperturbed compressibility
the object function
\eqref{eqn:theory_rel_fluctuations}, i.e.,
% \cite[\S35.2.1]{book:OlverNHMF2010}
% \cite[(1.38)]{book:Cohen2007},
% \cite[(4)]{article:LubbockPIEE1969}
\begin{equation}
 %--------------------------------------------------------------------------------------------------------------
 % Laplace transform of the relative spatial fluctuations in the unperturbed compressibility
 %--------------------------------------------------------------------------------------------------------------
  \Gamma( \hat{\vect{s}} )
%   =
%   \Gamma( \hat{s}_{x}, \hat{s}_{y}, \hat{s}_{z} )
%   =
%   \int_{ \Omega }
%     \gamma( \vect{r} )
%     e^{ - \dinprodr{ \hat{\vect{s}} }{ \vect{r} }{0} }
%   \text{d} \vect{r}
  =
  \int_{ \Omega }
    \gamma( \vect{r} )
    e^{ - ( \hat{s}_{x} x + \hat{s}_{y} y + \hat{s}_{z} z ) }
  \text{d} \vect{r}.
 \label{eqn:app_fdt_pw_LT_kappa}
\end{equation}

%%%%%%%%%%%%%%%%%%%%%%%%%%%%%%%%%%%%%%%%%%%%%%%%%%%%%%%%%%%%%%%%%%%%%%%%%%%%%%%%%%%%%%%%%%%%%%%%%%%%%%%%%%%%%%%%
% 3.) element height
%%%%%%%%%%%%%%%%%%%%%%%%%%%%%%%%%%%%%%%%%%%%%%%%%%%%%%%%%%%%%%%%%%%%%%%%%%%%%%%%%%%%%%%%%%%%%%%%%%%%%%%%%%%%%%%%
\subsection{Element Height}
% \label{subsec:theory_f_number_receive}
%---------------------------------------------------------------------------------------------------------------
% 1.) element height
%---------------------------------------------------------------------------------------------------------------
% a)
The element height
$h_{\text{e}}$ is
% 1.) incorporated
incorporated by
% 2.) integrating
integrating
% 3.) Born approximation of the scattered acoustic pressure field [ compressibility; Fourier domain; plane-wave decomposition; rigid baffle ]
the scattered acoustic pressure field
\eqref{eqn:theory_p_sc} over
% 4.) elevational interval
the elevational interval
$y \in [ - h_{\text{e}} / 2; h_{\text{e}} / 2 ]$.
% b)
%  integral in expression:
%
%    \int_{ - h_{\text{e}} / 2 }^{ h_{\text{e}} / 2 }
%      e^{ j k_{y} y }
%    \text{d} y
%    = [ k_{y} \neq 0 ]
%    \frac{
%      e^{ j k_{y} h_{\text{e}} / 2 }
%      -
%      e^{ -j k_{y} h_{\text{e}} / 2 }
%    }{
%      j k_{y}
%    }
%    = [ k_{y} \neq 0 ]
%    h_{\text{e}}
%    \frac{
%      \sin( k_{y} h_{\text{e}} / 2 )
%    }{
%      k_{y} h_{\text{e}} / 2
%    }
%    =
%    h_{\text{e}}
%    \sinc( k_{y} h_{\text{e}} / 2 )
This integral, denoted by
% 1.) \bar{p}^{(\text{sc})}( x )
$\bar{p}^{(\text{sc})}( x )$, equals
% 2.) inverse Fourier transform
the inverse Fourier transform
\begin{equation*}
 %--------------------------------------------------------------------------------------------------------------
 % complex-valued angular spatial frequencies [steered PWs]
 %--------------------------------------------------------------------------------------------------------------
  \bar{p}^{(\text{sc})}( x )
%   =
%   \int_{ - h_{\text{e}} / 2 }^{ h_{\text{e}} / 2 }
%     p^{(\text{sc})}( x, y )
%   \text{d} y
%   =
%   \frac{
%     j \munderbar{k}
%     p^{(\text{in})}( \vect{0} )
%   }{
%     ( 2 \pi )^{2}
%   }
%   \int_{ \opendisk{ \sigma } }
%     \frac{
%       \Gamma\left[ \vect{T}_{\vartheta}( \vect{k}_{\rho} ) \right]
%       e^{ j k_{x} x }
%     }{
%       Z_{0} / Z_{2}
%       +
%       \munderbar{k}_{z}( \vect{k}_{\rho} ) / \munderbar{k}
%     }
%     \int_{ - h_{\text{e}} / 2 }^{ h_{\text{e}} / 2 }
%       e^{ j k_{y} y }
%     \text{d} y
%   \text{d} \vect{k}_{\rho}
%   =
%   \frac{
%     j \munderbar{k}
%     p^{(\text{in})}( \vect{0} )
%     h_{\text{e}}
%   }{
%     ( 2 \pi )^{2}
%   }
%   \int_{ \opendisk{ \sigma } }
%     \frac{
%       \Gamma\left[ \vect{T}_{\vartheta}( \vect{k}_{\rho} ) \right]
%       e^{ j k_{x} x }
%     }{
%       Z_{0} / Z_{2}
%       +
%       \munderbar{k}_{z}( \vect{k}_{\rho} ) / \munderbar{k}
%     }
%     \sinc( k_{y} h_{\text{e}} / 2 )
%   \text{d} \vect{k}_{\rho}
%   =
%   \frac{
%     1
%   }{
%     2 \pi
%   }
%   \int_{ -\sigma }^{ \sigma }
%     \frac{
%       j \munderbar{k}
%       p^{(\text{in})}( \vect{0} )
%       h_{\text{e}}
%     }{
%       2 \pi
%     }
%     \int_{ - \sqrt{ \sigma^{2} - { k_{x} }^{2} } }^{ \sqrt{ \sigma^{2} - { k_{x} }^{2} } }
%       \frac{
%         \sinc( k_{y} h_{\text{e}} / 2 )
%         \Gamma\left[ \vect{T}_{\vartheta}( \vect{k}_{\rho} ) \right]
%       }{
%         Z_{0} / Z_{2}
%         +
%         \munderbar{k}_{z}( \vect{k}_{\rho} ) / \munderbar{k}
%       }
%     \text{d} k_{y}
%     e^{ j k_{x} x }
%   \text{d} k_{x}
  =
  \frac{
    1
  }{
    2 \pi
  }
  \int_{ -\sigma }^{ \sigma }
    \bar{P}^{(\text{sc})}( k_{x} )
    e^{ j k_{x} x }
  \text{d} k_{x},
\end{equation*}
where
% 5.) \bar{P}^{(\text{sc})}( k_{x} )
$\bar{P}^{(\text{sc})}( k_{x} )$ denotes
% 6.) lateral spectrum
the lateral spectrum
\begin{equation}
 %--------------------------------------------------------------------------------------------------------------
 % lateral spectrum of the scattered acoustic pressure field [ plane z = 0 ]
 %--------------------------------------------------------------------------------------------------------------
  \bar{P}^{(\text{sc})}( k_{x} )
%   =
%   \frac{
%     j \munderbar{k}
%     p^{(\text{in})}( \vect{0} )
%     h_{\text{e}}
%   }{
%     2 \pi
%   }
%   \int_{ - \sqrt{ \sigma^{2} - { k_{x} }^{2} } }^{ \sqrt{ \sigma^{2} - { k_{x} }^{2} } }
%     \frac{
%       \sinc( k_{y} h_{\text{e}} / 2 )
%       \Gamma\left[ \vect{T}_{\vartheta}( \vect{k}_{\rho} ) \right]
%     }{
%       Z_{0} / Z_{2}
%       +
%       \munderbar{k}_{z}( \vect{k}_{\rho} ) / \munderbar{k}
%     }
%   \text{d} k_{y}
  =
  C_{0}
  \int_{ k_{y, \text{lb}} }^{ k_{y, \text{ub}} }
    \frac{
      \sinc( k_{y} h_{\text{e}} / 2 )
      \Gamma\left[ \vect{T}_{\vartheta}( \vect{k}_{\rho} ) \right]
    }{
      \munderbar{k}_{z}( \vect{k}_{\rho} )
    }
  \text{d} k_{y}
 \label{eqn:theory_lateral_spectrum}
\end{equation}
with
% 7.)
$C_{0} = j \munderbar{k}^{2} p^{(\text{in})}( \vect{0} ) h_{\text{e}} / ( 2 \pi )$,
% 8.) integration limits
the integration limits
% 8.a) k_{y, \text{lb}} = - \sqrt{ \sigma^{2} - { k_{x} }^{2} }
$k_{y, \text{lb}} = - \sqrt{ \sigma^{2} - { k_{x} }^{2} }$ and
% 8.b) k_{y, \text{ub}}
$k_{y, \text{ub}} = - k_{y, \text{lb}}$, and
% 9.) sinc function
the sinc function
$\sinc( x ) = \sin( x ) / x$.

%%%%%%%%%%%%%%%%%%%%%%%%%%%%%%%%%%%%%%%%%%%%%%%%%%%%%%%%%%%%%%%%%%%%%%%%%%%%%%%%%%%%%%%%%%%%%%%%%%%%%%%%%%%%%%%%
% 4.) spatial discretization
%%%%%%%%%%%%%%%%%%%%%%%%%%%%%%%%%%%%%%%%%%%%%%%%%%%%%%%%%%%%%%%%%%%%%%%%%%%%%%%%%%%%%%%%%%%%%%%%%%%%%%%%%%%%%%%%
\subsection{Spatial Discretization}
% \label{subsec:theory_f_number_receive}
%---------------------------------------------------------------------------------------------------------------
% 1.) spatial discretization
%---------------------------------------------------------------------------------------------------------------
% a)
Both
% 1.) Laplace transform of the relative spatial fluctuations in the unperturbed compressibility
the \name{Laplace} transform
\eqref{eqn:app_fdt_pw_LT_kappa} and
% 2.) lateral spectrum of the scattered acoustic pressure field [ plane z = 0 ]
the lateral spectrum
\eqref{eqn:theory_lateral_spectrum} are
% 3.) discretized
discretized by
% 4.) approximating
approximating
% 5.) relative spatial fluctuations in the unperturbed adiabatic compressibility
the object function
\eqref{eqn:theory_rel_fluctuations} inside
% 6.) field of view (FOV)
the \ac{FOV}
$\Omega_{\text{FOV}}$ by
% 7.) Fourier sum
a Fourier sum.
% b)
This approximation is
% 1.) valid
valid because both
% 2.) field of view (FOV)
the \ac{FOV}
$\Omega_{\text{FOV}}$ and
% 3.) passbands
the passbands are
% 4.) bounded
bounded
[see \cref{%
  fig:theory_measurement_setup,%
  fig:theory_passbands_pw_broadband_sequence%
}].
% c)
The \ac{FOV}
$\Omega_{\text{FOV}}$ is
% 1.) box-shaped
box-shaped and
% 2.) transversally centered
transversally centered in
% 3.) front
front of
% 4.) array
the array
(i.e.,
  % 5.)
  $x \in [ - L_{x} / 2; L_{x} / 2 ]$,
  % 6.)
  $z \in [ z_{\text{lb}}; z_{\text{lb}} + L_{z} ]$ with
  % 7.) edge lengths
  the edge lengths
  % 7.a) L_{x}
  $L_{x}$ and
  % 7.b) L_{z}
  $L_{z}$,
  % 8.) axial lower bound
  the axial lower bound
  $z_{\text{lb}} > 0$, and
  % 9.) elevational slice thickness
  the elevational slice thickness
  % 10.) \Delta y
  $\Delta y$%
).
% d)
The Fourier sum, hence, reads
\begin{equation}
 %--------------------------------------------------------------------------------------------------------------
 % relative spatial fluctuations in the unperturbed adiabatic compressibility [ Fourier sum ]
 %--------------------------------------------------------------------------------------------------------------
  \gamma( \vect{r} )
  \approx
  \Delta y
  \delta( y )
  \sum_{ \nu = \nu_{\text{lb}} }^{ \nu_{\text{ub}} }
    \sum_{ \mu = \mu_{\text{lb}} }^{ \mu_{\text{ub}} }
      \xi_{ \nu, \mu }
      e^{ j ( \hat{k}_{x, \nu} x + \hat{k}_{z, \mu} z ) },
  \label{eqn:theory_Fourier_sum}
\end{equation}
where
% 1.) \delta
$\delta$ denotes
% 2.) Dirac delta function
the Dirac delta function,
% 3.) \xi_{ \nu, \mu }
$\xi_{ \nu, \mu }$ are
% 4.) unknown coefficients
unknown coefficients, and
% 5.a) \hat{k}_{x, \nu}
$\hat{k}_{x, \nu} = 2 \pi \nu / L_{x}$ and
% 5.b) \hat{k}_{z, \mu}
$\hat{k}_{z, \mu} = 2 \pi \mu / L_{z}$ are
% 6.a) lateral angular frequencies
the lateral and
% 6.b) axial angular frequencies
axial angular frequencies,
respectively.
% e)
The index bounds
% 1.a) \nu_{\text{lb}}
$\nu_{\text{lb}}$,
% 1.b) \nu_{\text{ub}}
$\nu_{\text{ub}}$,
% 2.a) \mu_{\text{lb}}
$\mu_{\text{lb}}$, and
% 2.b) \mu_{\text{ub}}
$\mu_{\text{ub}}$ depend on
% 3.) passbands
the passbands and set
% 4.) required number
the number of
% 5.) coefficients
coefficients
$N_{\text{c}} = ( \nu_{\text{ub}} - \nu_{\text{lb}} + 1 ) ( \mu_{\text{ub}} - \mu_{\text{lb}} + 1 )$.

%%%%%%%%%%%%%%%%%%%%%%%%%%%%%%%%%%%%%%%%%%%%%%%%%%%%%%%%%%%%%%%%%%%%%%%%%%%%%%%%%%%%%%%%%%%%%%%%%%%%%%%%%%%%%%%%
% 1.) laplace transform
%%%%%%%%%%%%%%%%%%%%%%%%%%%%%%%%%%%%%%%%%%%%%%%%%%%%%%%%%%%%%%%%%%%%%%%%%%%%%%%%%%%%%%%%%%%%%%%%%%%%%%%%%%%%%%%%
\subsubsection{Laplace Transform}

The Laplace transform
\eqref{eqn:app_fdt_pw_LT_kappa}, inserting
% 1.) relative spatial fluctuations in the unperturbed adiabatic compressibility [ Fourier sum ]
the Fourier sum
\eqref{eqn:theory_Fourier_sum}, yields
\begin{equation}
\begin{split}
 %--------------------------------------------------------------------------------------------------------------
 % Laplace transform of the relative spatial fluctuations in the unperturbed compressibility
 %--------------------------------------------------------------------------------------------------------------
  \Gamma( \hat{\vect{s}} )
  =\,
& \Delta V
  \sum_{ \nu = \nu_{\text{lb}} }^{ \nu_{\text{ub}} }
    \sum_{ \mu = \mu_{\text{lb}} }^{ \mu_{\text{ub}} }
      \xi_{ \nu, \mu }
      \sinhc\Bigl[ \bigl( \hat{s}_{x} - j \hat{k}_{x, \nu} \bigr) \frac{ L_{x} }{ 2 } \Bigr]\\
&     \times
      \sinhc\Bigl[ \bigl( \hat{s}_{z} - j \hat{k}_{z, \mu} \bigr) \frac{ L_{z} }{ 2 } \Bigr]
      e^{ - ( \hat{s}_{z} - j \hat{k}_{z, \mu} ) z_{\text{c}} },
\end{split}
\label{eqn:theory_LT_discretized}
\end{equation}
where
% 1.) \Delta V
$\Delta V = L_{x} \Delta y L_{z}$ and
% 2.) z_{\text{c}}
$z_{\text{c}} = z_{\text{lb}} + L_{z} / 2$ denote
% 3.) volume
the volume and
% 4.) axial center
the axial center of
% 5.) field of view (FOV)
the \ac{FOV},
respectively,
% 6.) \sinhc
$\sinhc$ is
% 7.) hyperbolic version
the hyperbolic version of
% 8.) cardinal sine function
the sinc function
\begin{equation*}
 %--------------------------------------------------------------------------------------------------------------
 % hyperbolic sinc function
 %--------------------------------------------------------------------------------------------------------------
  \sinhc( \mathcal{Z} )
  =
  \begin{cases}
   %--------------------------------------------------------------------------------------------------------------
   % nonzero
   %--------------------------------------------------------------------------------------------------------------
    \sinh( \mathcal{Z} ) / \mathcal{Z}
  & \text{for}\, \mathcal{Z} \neq 0,\\
   %--------------------------------------------------------------------------------------------------------------
   % zero
   %--------------------------------------------------------------------------------------------------------------
    1
  & \text{for}\, \mathcal{Z} = 0,
  \end{cases}
 \label{eqn:theory_discretization_sinhc}
\end{equation*}
and
% 9.) \sinh
$\sinh( \mathcal{Z} ) = ( e^{ \mathcal{Z} } - e^{ - \mathcal{Z} } ) / 2$ is
% 10.) hyperbolic sine
the hyperbolic sine.

\subsubsection{Lateral Spectrum}
%---------------------------------------------------------------------------------------------------------------
% 1.) spatial discretization of the lateral spectrum
%---------------------------------------------------------------------------------------------------------------
% a)
The lateral spectrum
% 1.) lateral spectrum of the scattered acoustic pressure field [ plane z = 0 ]
\eqref{eqn:theory_lateral_spectrum}, inserting
% 2.) discretized Laplace transform
the discretized Laplace transform
\eqref{eqn:theory_LT_discretized}, becomes
\begin{subequations}
\label{eqn:theory_lateral_spectrum_discretized}
\begin{equation}
 %--------------------------------------------------------------------------------------------------------------
 % lateral spectrum of the scattered acoustic pressure field [ plane z = 0, discretized ]
 %--------------------------------------------------------------------------------------------------------------
  \bar{P}^{(\text{sc})}( k_{x} )
  =
  \sum_{ \nu = \nu_{\text{lb}} }^{ \nu_{\text{ub}} }
    \sum_{ \mu = \mu_{\text{lb}} }^{ \mu_{\text{ub}} }
      a_{\nu}( k_{x} )
      b_{\mu}( k_{x} )
      \xi_{ \nu, \mu },
 \label{eqn:theory_lateral_spectrum_discretized_expression}
\end{equation}
where
% 3.)
$a_{\nu}( k_{x} )$ and
% 4.)
$b_{\mu}( k_{x} )$ denote
% 5.)
the coefficients
\begin{align}
 %--------------------------------------------------------------------------------------------------------------
 % lateral coefficient
 %--------------------------------------------------------------------------------------------------------------
  a_{\nu}( k_{x} )
  &=
  D_{0}
  \sinhc\Bigl[ \bigl( T_{\vartheta, x}( k_{x} ) - j \hat{k}_{x, \nu} \bigr) \frac{ L_{x} }{ 2 } \Bigr],
 \label{eqn:theory_lateral_spectrum_discretized_coef_lateral}\\
 %--------------------------------------------------------------------------------------------------------------
 % lateral spectrum of the scattered acoustic pressure field [ plane z = 0, discretized, axial coefficient ]
 %--------------------------------------------------------------------------------------------------------------
  b_{\mu}( k_{x} )
%   &=
%   \begin{aligned}[t]
%     \int_{ k_{y, \text{lb}} }^{ k_{y, \text{ub}} }
%     & \frac{
%         \sinc( k_{y} h_{\text{e}} / 2 )
%         e^{ - ( T_{\vartheta, z}( \vect{k}_{\rho} ) - j \hat{k}_{z, \mu} ) z_{\text{c}} }
%       }{
%         Z_{0} / Z_{2}
%         +
%         \munderbar{k}_{z}( \vect{k}_{\rho} ) / \munderbar{k}
%       }\\
%     & \times
%       \sinhc\Bigl[ \bigl( T_{\vartheta, z}( \vect{k}_{\rho} ) - j \hat{k}_{z, \mu} \bigr) \frac{ L_{z} }{ 2 } \Bigr]
%     \text{d} k_{y}.
%   \end{aligned}
  &=
  \begin{aligned}[t]
  & \int_{ k_{y, \text{lb}} }^{ k_{y, \text{ub}} }
      \frac{
        e^{ - [ T_{\vartheta, z}( \vect{k}_{\rho} ) - j \hat{k}_{z, \mu} ] z_{\text{c}} }
      }{
        \munderbar{k}_{z}( \vect{k}_{\rho} )
      }
      \sinc\Bigl( k_{y} \frac{ h_{\text{e}} }{ 2 } \Bigr)\\
  &   \times
      \sinhc\Bigl[ \bigl( T_{\vartheta, z}( \vect{k}_{\rho} ) - j \hat{k}_{z, \mu} \bigr) \frac{ L_{z} }{ 2 } \Bigr]
      \text{d} k_{y}
  \end{aligned}
 \label{eqn:theory_lateral_spectrum_discretized_coef_axial}
\end{align}
\end{subequations}
with
% 6.)
$D_{0} = C_{0} \Delta V$.

\subsection{Image Formation by Inversion}
% \label{subsec:theory_inversion}
%---------------------------------------------------------------------------------------------------------------
% image formation by inversion
%---------------------------------------------------------------------------------------------------------------
% a)
Images are
% 1.) formed
formed by
% 2.) computing
computing
% 3.) N_{\text{c}} coefficients
the coefficients
$\xi_{ \nu, \mu }$ of
% 4.) relative spatial fluctuations in the unperturbed adiabatic compressibility [ Fourier sum ]
the \name{Fourier} sum
\eqref{eqn:theory_Fourier_sum} from
% 5.) lateral spectrum of the scattered acoustic pressure field [ plane z = 0, discretized ]
the discretized lateral spectrum
\eqref{eqn:theory_lateral_spectrum_discretized}.
% % b)
% This spectrum can be
% % 1.) computed
% computed from
% % 2.) recorded echo signals
% the recorded echo signals.
% b)
This spectrum, considering
% 1.) N_{\text{r}} discrete combinations
$N_{\text{r}}$ discrete combinations of
% 2.) temporal frequency
temporal
$f$ and
% 3.) angular lateral frequency
angular lateral frequency
$k_{x}$ across
% 4.) multiple pulse-echo measurements
multiple pulse-echo measurements, defines
% 5.) linear equation system
a linear equation system.
% c)
The
$N_{\text{r}} \times N_{\text{c}}$
system matrix
$\mat{A}$ contains
% 1.) products
the products of
% 2.) coefficients
the coefficients
% 2.a) lateral coefficient
\eqref{eqn:theory_lateral_spectrum_discretized_coef_lateral} and
% 2.b) axial coefficient
\eqref{eqn:theory_lateral_spectrum_discretized_coef_axial} and is
% 3.) ill-conditioned
ill-conditioned.
% d)
The solution, hence, is
% 1.) regularized
regularized by
% 2.) solving
solving
% 3.) optimization problem
the optimization problem
\begin{equation}
 %--------------------------------------------------------------------------------------------------------------
 % l2-minimization problem
 %--------------------------------------------------------------------------------------------------------------
  \hat{\vectsym{\xi}}
  =
  \underset{ \tilde{ \vectsym{\xi} } \in \C^{ N_{\text{c}} } }{ \arg\min }
  \dnorm{ \tilde{ \vectsym{\xi} } }{2}{1}
  \quad
  \text{such that}
  \quad
  \dnorm{ \vect{b} - \mat{A} \tilde{ \vectsym{\xi} } }{2}{1}
  <
  \epsilon,
 \label{eqn:theory_inversion}
\end{equation}
where
% 4.) \hat{\vectsym{\xi}}
$\hat{\vectsym{\xi}}$ estimates
% 5.) Fourier coefficients
the \name{Fourier} coefficients,
% 6.) \vect{b}
$\vect{b}$ contains
% 7.) measured samples
measured samples of
% 8.) lateral spectrum of the scattered acoustic pressure field [ plane z = 0, discretized ]
the lateral spectrum
\eqref{eqn:theory_lateral_spectrum_discretized} and
% 9.) \epsilon
$\epsilon > 0$ accounts for
% 10.) model mismatches
model mismatches and
% 11.) noise
noise.

% \frac{
%   e^{ a ( x + j y) }
%   -
%   e^{ -a ( x + j y) }
% }{
%   x + j y
% }
% =
% \frac{
%   \cos( a y ) [ e^{ a x } - e^{ -a x } ]
%   + j \sin( a y ) [ e^{ a x } + e^{ -a x } ]
% }{
%   x + j y
% }
% =
% 2
% \frac{
%   \cos( a y ) \sinh( a x )
%   +
%   j \sin( a y ) \cosh( a x )
% }{
%   x + j y
% }
% =
% 2
% \frac{
%   \sqrt{ \cos^{2}( a y ) \sinh^{2}( a x ) + \sin^{2}( a y ) \cosh^{2}( a x ) }
% }{
%   \sqrt{ x^{2} + y^{2} }
% }

% $\setsymbol{M} = \left\{ \hat{\vect{k}} \in \R^{2}: \beta \uvect{\vartheta} + \trans{ ( k_{x}, \munderbar{k}_{z}( k_{x} ) ) }, \tabs{ k_{x} } < \beta, \beta_{\text{lb}} \leq \beta \leq \beta_{\text{ub}} \right\}$
%

%%%%%%%%%%%%%%%%%%%%%%%%%%%%%%%%%%%%%%%%%%%%%%%%%%%%%%%%%%%%%%%%%%%%%%%%%%%%%%%%%%%%%%%%%%%%%%%%%%%%%%%%%%%%%%%%
% 4.) methods
%%%%%%%%%%%%%%%%%%%%%%%%%%%%%%%%%%%%%%%%%%%%%%%%%%%%%%%%%%%%%%%%%%%%%%%%%%%%%%%%%%%%%%%%%%%%%%%%%%%%%%%%%%%%%%%%
\section{Methods}
\label{sec:methods}
%---------------------------------------------------------------------------------------------------------------
% 1.) methods
%---------------------------------------------------------------------------------------------------------------
% a)
The proposed \name{Laplace}-domain beamforming was
% 1.) compared
compared to
% 2.) established Fourier-domain methods
\name{Fourier}-domain methods in
% 3.) experiment
an experiment with
% 4.) commercial model 040 multi-purpose multi-tissue ultrasound phantom
a commercial multi-tissue phantom%
\footnote{% (CIRS)
  Computerized Imaging Reference Systems, Inc., Norfolk, VA, USA%
  \label{ftnote:manufacturer_cirs}
}
[%
  model: 040;
  average speed of sound: $\bar{c} = \qty{1538.75}{\meter\per\second}$;
  absorption: \qty{0.49}{\deci\bel\per\mega\hertz\per\centi\meter}%
].
% b)
A SonixTouch Research system%
\footnote{%
  Analogic Corporation, Sonix Design Center, Richmond, BC, Canada
  \label{ftnote:manufacturer_analogic}
} with
% 1.) linear array
% L14-5/38  & 128 & 7.2 & 70 & 279.8 & 25 & 304.8 & 4.0 & 16 & 0.519803852\\% \lambda_{\text{lb}} = 158.436213992 um
% L14-5/38 [ w = 279.8 um | k = 25 um | \Delta x_{\text{el}} = 304.8 um ]
% 1.a.1) total number of identical array elements (geometric parameter of the linear transducer array)
% 1.a.2) number of active faces along the r_{1}-axis (geometric parameter of the linear transducer array)
% $N_{\text{el}} = N_{\text{el}, 1} = 128$
% 1.b) width of the identical kerfs separating the active faces along the r_{1}-axis (geometric parameter of the linear transducer array)
% $k_{\text{el}, 1} = \qty{25}{\micro\meter}$
% 2.a.1) constant spacing between the centers of the adjacent faces along the r_{1}-axis (element pitch, geometric parameter of the linear transducer array)
% 2.a.2) identical width of the active faces along the r_{1}-axis (geometric parameter of the linear transducer array)
% $w_{\text{el}, 1} = \qty{279.8}{\micro\meter}$
% 3.b) fractional bandwidth (electromechanical parameter of the instrumentation)
%   $B_{h, \text{frac}} = 0.7$
% 3.a) center frequency (electromechanical parameter of the instrumentation)
%   $f_{\text{c}} = \omega_{\text{c}} / ( 2 \pi ) = \qty{4}{\mega\hertz}$
a linear array
(%
  % 1.a) model
  model: L14-5/38;
  % 1.b) number of elements
  $N_{\text{e}} = \num{128}$,
  % 1.c) pitch
  $p_{\text{e}} = \qty{304.8}{\micro\meter}$,
  % 1.d) element height
  $h_{\text{e}} = \qty{4}{\milli\meter}$%
) acquired and stored
% 2.) plane-wave scan
a plane-wave scan for
% 3.) offline processing
offline processing.
% c)
The steering angles were
% 1.) 85°, 90°, and 95°
$\vartheta \in \{ \qty{85}{\degree}, \qty{90}{\degree}, \qty{95}{\degree} \}$.
% d)
The excitation voltage was
% 1.) single cycle
a single cycle at
% 2.) 4 MHz
\qty{4}{\mega\hertz}.
% e)
% bandlimits:
%   2.25 MHz - 6.75 MHz
% => k_{lb} = 2 \pi 2.25 MHz / 1538.75 m / s \approx 9180 rad / m
% => k_{ub} = 2 \pi 6.75 MHz / 1538.75 m / s \approx 27540 rad / m
% The lower and
% 1.) upper frequency bound
% upper frequency bounds were
The frequency bounds were
% 2.) lower cutoff frequency (lower frequency bound)
$f_{\text{lb}} = \qty{2.25}{\mega\hertz}$ and
% 3.) upper cutoff frequency (upper frequency bound)
$f_{\text{ub}} = \qty{6.75}{\mega\hertz}$. %,
% respectively.
% f)
% MATLAB R2021a
% (The MathWorks, Inc., Natick, MA, USA) enabled
% % 1.)
% all numerical calculations.

%%%%%%%%%%%%%%%%%%%%%%%%%%%%%%%%%%%%%%%%%%%%%%%%%%%%%%%%%%%%%%%%%%%%%%%%%%%%%%%%%%%%%%%%%%%%%%%%%%%%%%%%%%%%%%%%
% 1.) image formation
%%%%%%%%%%%%%%%%%%%%%%%%%%%%%%%%%%%%%%%%%%%%%%%%%%%%%%%%%%%%%%%%%%%%%%%%%%%%%%%%%%%%%%%%%%%%%%%%%%%%%%%%%%%%%%%%
\subsection{Image Formation}
\label{subsec:methods_image_formation}
%---------------------------------------------------------------------------------------------------------------
% 1.) image formation
%---------------------------------------------------------------------------------------------------------------
% a)
Images were
% 1.) formed
formed by
% 2.) filtered backpropagation (FBP)
\ac{FBP}
\cite{proc:SchiffnerAI2012} and by
% 3.) solving
solving
% 4.) l2-minimization problem
the optimization problem
\eqref{eqn:theory_inversion}.
% b)
This problem was
% 1.) solved iteratively
\TODO{algorithm?}
solved iteratively across
% 2.) four combinations
four combinations of
% 3.) wavenumber
wavenumber
\eqref{eqn:intro_wavenumber} and
% 4.) spatial dimensionality
spatial dimensionality, representing
% 5.) Laplace-domain methods
two \name{Laplace}-domain and
% 6.) Fourier-domain methods
two \name{Fourier}-domain methods.
% c)
The \name{Laplace}-domain methods used
the complex time-causal wavenumber
\cite[Eq. (3.98)]{book:Cobbold2006}.
% \cite[Table II]{article:WatersITUFFC2005},
% \cite[Eqs. (6) and (7)]{article:CobboldJASA2004}.
% d)
The \name{Fourier}-domain methods, in contrast, used
% 1.) standard real wavenumber
the standard real wavenumber
(i.e.,
  % 2.) absorption coefficient
  $\alpha( f ) = 0$ and
  % 3.) phase coefficient
  $\beta( f ) = 2 \pi f / \bar{c}$ for
  % 4.) all frequencies
  all $f$%
).
% e)
% The \name{Laplace} transform
% % 1.) Laplace transform of the relative spatial fluctuations in the unperturbed compressibility
% \eqref{eqn:app_fdt_pw_LT_kappa}, in this case, reduced to
% % 2.) Fourier transform
% a \name{Fourier} transform.
% f)
Both methods were
% 1.) evaluated
evaluated in
% 2.) two-dimensional space
two- and
% 3.) three-dimensional space
three-dimensional space.
% g)
% FOV:
%   7.8 cm x 4.37 cm x 1 mm
The \ac{FOV}
$\Omega_{\text{FOV}}$ had
% 1.) parameters
the parameters
% 2.a) lateral edge length
$L_{x} = \qty{7.8}{\centi\meter}$,
% 2.a) axial edge length
$L_{z} = \qty{4.37}{\centi\meter}$,
% 3.) elevational slice thickness
$\Delta y = \qty{1}{\milli\meter}$, and
% 4.) z_{\text{lb}}
$z_{\text{lb}} = \qty{5}{\milli\meter}$.
These settings resulted in
% 1.) system matrix
a system matrix
$\mat{A}$ of
% 2.)
% number of rows:
%   3 * 256 * 370 = 284160 | propagable: 283197
% MATLAB:
% N_theta * size( mat_k_bp_k_x_propagable, 1 ) = 283197
%
% number of cols:
%   743 * 320 = 237760
% MATLAB:
% N_coef_x * N_coef_z = 237760
size $\num{283197} \times \num{237760}$.
% j)
% MATLAB:
% nnz( A ) * 1e2 / numel( A ) = 0.0261
%
% whos A
% Name   Size             Bytes      Class     Attributes     
% A      283197 x 237760  424280800  double    sparse, complex
%
% => 424280800 Bytes = 404.625701904 MiBytes = 424.2808 MBytes
However,
% 1.) up to 0.3%
% lossy, 3d:     nnz( A ) = 199525771 (0.296327227%) | whos A: A 283197 x 237760 4790520592 double sparse, complex | 4.7905 GByte
% lossless, 3d:  nnz( A ) = 118201515 (0.175547885%) | whos A: A 283191 x 237760 2838738448 double sparse, complex | 2.8387 GByte
% lossy, 2d:     nnz( A ) = 205280664 (0.304874151%) | whos A: A 283197 x 237760 4928638024 double sparse, complex | 4.9286 GByte
% lossless, 2d:  nnz( A ) = 117989269 (0.175232667%) | whos A: A 283191 x 237760 2833644544 double sparse, complex | 2.8336 GByte
only up to \qty{0.3}{\percent} of
% 2.) entries
its entries were
% 3.) relevant
relevant and required
% 4.) storage
storage
(memory footprint: \qty{4.9}{\giga\byte}).
% k)
Consequently,
% 1.) matrix-vector products
the matrix-vector products in
% 2.) each iteration
each iteration were
% 3.) computationally efficient
computationally efficient.
% l)
% software:LaplaceInversion: Laplace-domain beamforming for ultrafast plane-wave imaging
% software:FBP: Filtered backpropagation for coherent plane-wave compounding
To support
% 1.) reproducibility
reproducibility,
% 2.) Matlab source code
the \name{Matlab}%
\footnote{%
  The MathWorks, Inc., Natick, MA, USA
  \label{ftnote:manufacturer_mathworks}
} source code is
% 3.) publicly available
publicly available
\cite{software:LaplaceInversion}.

\subsection{Evaluation Metrics}
\label{subsec:methods_post_processing}
%---------------------------------------------------------------------------------------------------------------
% 1.) image quality metrics
%---------------------------------------------------------------------------------------------------------------
% c) interpolation simplified comparisons by the mean SSIM indices and the relative RMSEs
The lateral and axial \acp{FWHM} of
% 1.) all wires
all wires and
% 2.) generalized contrast-to-noise ratios (gCNRs)
% article:Rodriguez-MolaresITUFFC2020: The Generalized Contrast-to-Noise Ratio: A Formal Definition for Lesion Detectability
the \acp{GCNR}
\cite{article:Rodriguez-MolaresITUFFC2020} of
% 3.) anechoic regions
anechoic regions were
% 4.) computed
computed.

%%%%%%%%%%%%%%%%%%%%%%%%%%%%%%%%%%%%%%%%%%%%%%%%%%%%%%%%%%%%%%%%%%%%%%%%%%%%%%%%%%%%%%%%%%%%%%%%%%%%%%%%%%%%%%%%
% 5.) results
%%%%%%%%%%%%%%%%%%%%%%%%%%%%%%%%%%%%%%%%%%%%%%%%%%%%%%%%%%%%%%%%%%%%%%%%%%%%%%%%%%%%%%%%%%%%%%%%%%%%%%%%%%%%%%%%
\section{Results}
% \label{sec:results}
%%%%%%%%%%%%%%%%%%%%%%%%%%%%%%%%%%%%%%%%%%%%%%%%%%%%%%%%%%%%%%%%%%%%%%%%%%%%%%%%%%%%%%%%%%%%%%%%%%%%%%%%%%%%%%%%
% table: image quality metrics
%%%%%%%%%%%%%%%%%%%%%%%%%%%%%%%%%%%%%%%%%%%%%%%%%%%%%%%%%%%%%%%%%%%%%%%%%%%%%%%%%%%%%%%%%%%%%%%%%%%%%%%%%%%%%%%%
\begin{table}[tb!]
 \centering
 \caption{%
  % a)
  \acsp{FWHM} of
  % 1.) wires
  the wires and
  % 2.) generalized contrast-to-noise ratios (gCNRs)
  \acsp{GCNR} of
  % 3.) anechoic regions
  the anechoic regions.
  % b)
  The proposed \name{Laplace}-domain beamforming outperformed
  % 1.) filtered backpropagation (FBP)
  \ac{FBP} in
  % 2.) terms
  terms of
  % 3.) spatial resolution
  spatial resolution and
  % 4.) contrast
  contrast.
 }
 \label{tab:metrics}
 \small
 \setlength\tabcolsep{5pt} % reduce horizontal whitespace (default value: 6pt)
 \begin{tabular}{%
 %--------------------------------------------------------------------------------------------------------------
 % columns
 %--------------------------------------------------------------------------------------------------------------
  @{}%
  l%                                                                                              01.) tissue losses
  l%                                                                                              02.) number of spatial dimensions
  S[table-format=3.0,table-number-alignment = right,table-auto-round,separate-uncertainty=true]%  03.) median lateral FWHM (um)
  S[table-format=3.0,table-number-alignment = right,table-auto-round,separate-uncertainty=true]%  04.) IQR (um)
  S[table-format=3.0,table-number-alignment = right,table-auto-round,separate-uncertainty=true]%  05.) median axial FWHM (um)
  S[table-format=3.0,table-number-alignment = right,table-auto-round,separate-uncertainty=true]%  06.) IQR (um)
  S[table-format=2.0,table-number-alignment = right,table-auto-round]%                            07.) gCNR small region (%)
  S[table-format=2.0,table-number-alignment = right,table-auto-round]%                            08.) gCNR large region (%)
  @{}%
 }
 \toprule
 %--------------------------------------------------------------------------------------------------------------
 % headings
 %--------------------------------------------------------------------------------------------------------------
  \multicolumn{1}{@{}H}{\multirow{3}{*}{Tissue losses}} &
  \multicolumn{1}{H}{\multirow{3}{*}{\begin{tabular}[c]{@{}c@{}}Number of\\spatial\\dimensions\end{tabular}}} &
  \multicolumn{2}{H}{Lateral \acsp{FWHM}} &
  \multicolumn{2}{H}{Axial \acsp{FWHM}} &
  \multicolumn{2}{H@{}}{\acs{GCNR}}\\
  \cmidrule(lr){3-4}\cmidrule(lr){5-6}\cmidrule(l){7-8}
  &
  &
  \multicolumn{1}{H}{Median} &
  \multicolumn{1}{H}{\acs{IQR}} &
  \multicolumn{1}{H}{Median} &
  \multicolumn{1}{H}{\acs{IQR}} &
  \multicolumn{1}{H}{Small} &
  \multicolumn{1}{H@{}}{Large}\\
  &
  & % \multicolumn{1}{H}{(1)}
  \multicolumn{1}{H}{(\unit{\micro\meter})} &
  \multicolumn{1}{H}{(\unit{\micro\meter})} &
  \multicolumn{1}{H}{(\unit{\micro\meter})} &
  \multicolumn{1}{H}{(\unit{\micro\meter})} &
  \multicolumn{1}{H}{(\unit{\percent})} &
  \multicolumn{1}{H@{}}{(\unit{\percent})}\\
  \cmidrule(r){1-1}\cmidrule(lr){2-2}\cmidrule(lr){3-3}\cmidrule(lr){4-4}\cmidrule(lr){5-5}\cmidrule(lr){6-6}\cmidrule(lr){7-7}\cmidrule(l){8-8}
 \addlinespace
 %--------------------------------------------------------------------------------------------------------------
 % content
 %--------------------------------------------------------------------------------------------------------------
 % filtered backpropagation (FBP)
 \multicolumn{1}{@{}l}{\scriptsize \acs{FBP}}\\
 \cmidrule(r){1-1}
  Lossless & Two & 533.4000 & 154.3050 & 455.2950 & 45.7200 & 80.17 & 63.95\\

 \addlinespace
 % optimization
 \multicolumn{1}{@{}l}{\scriptsize Optimization}\\
 \cmidrule(r){1-1}
  Lossless & Two & 489.6491 & 152.2769 & 465.5790 & 18.5093 & 84.21 & 65.13\\
Lossless & Proposed & 492.2746 & 150.9642 & 467.0028 & 24.2044 & 84.19 & 65.93\\
Proposed & Two & 498.8382 & 148.3387 & 455.6125 & 41.2899 & 84.80 & 71.47\\
Proposed & Proposed & 498.8382 & 148.3387 & 405.7799 & 46.9850 & 86.98 & 72.57\\

 \addlinespace
 \bottomrule
 \end{tabular}
\end{table}

%%%%%%%%%%%%%%%%%%%%%%%%%%%%%%%%%%%%%%%%%%%%%%%%%%%%%%%%%%%%%%%%%%%%%%%%%%%%%%%%%%%%%%%%%%%%%%%%%%%%%%%%%%%%%%%%
% images: CIRS 040 [ dynamic apertures ]
%%%%%%%%%%%%%%%%%%%%%%%%%%%%%%%%%%%%%%%%%%%%%%%%%%%%%%%%%%%%%%%%%%%%%%%%%%%%%%%%%%%%%%%%%%%%%%%%%%%%%%%%%%%%%%%%
%
\begin{figure}[t!]
 \centering%
  \input{results/figures/latex/results_experiments_cirs_040.tex}
 \vspace*{-5mm}
 \caption{}
 \label{fig:V}
\end{figure}
C
{% a) images of the multi-tissue phantom
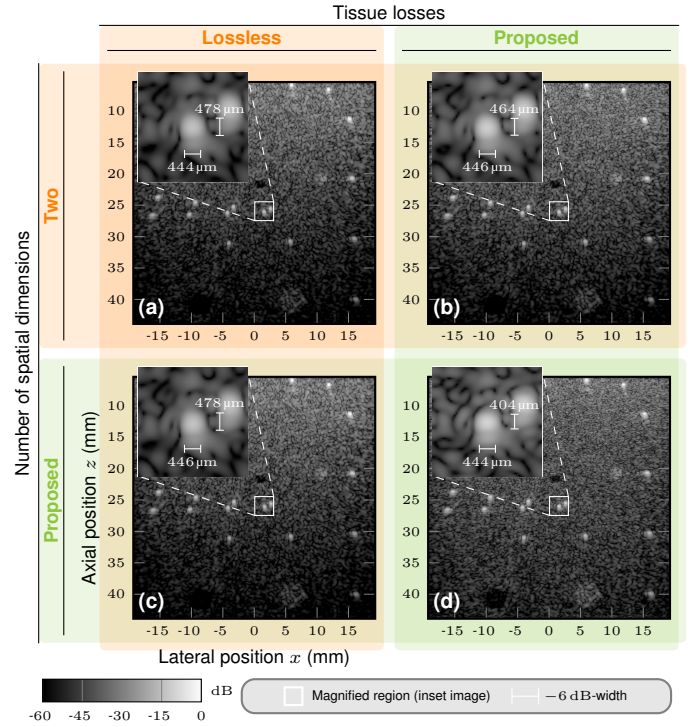
 Images of
 % 1.) commercial multi-tissue phantom
 the multi-tissue phantom.
 % b)
 The proposed \name{Laplace}-domain beamforming, which accounted for
 % 1.) power-law absorption
 power-law absorption and
 % 2.) three spatial dimensions
 three spatial dimensions, outperformed
 % 3.) all other investigated solutions
 all other investigated solutions of
 % 4.) l2-minimization problem
 the optimization problem
 \eqref{eqn:theory_inversion} in terms of
 % 5.) axial resolution
 axial resolution and
 % 6.) contrast
 contrast.
 % c)
 The lateral resolution was
 % 1.) comparable
 comparable.
 % d)
 The images show
 % 1.) relative spatial fluctuations in the unperturbed adiabatic compressibility [ Fourier sum ]
 the \name{Fourier} sum
 \eqref{eqn:theory_Fourier_sum} in
 % 2.) xz-plane
 the $xz$-plane
 (i.e., $y = 0$) for
 % 3.) all investigated combinations
 all investigated combinations of
 % 4.) complex-valued wavenumber
 wavenumber
 \eqref{eqn:intro_wavenumber} and
 % 5.) spatial dimensionality
 spatial dimensionality
 (%
   shown as
   % 6.) columns
   columns and
   % 7.) rows
   rows,
   respectively%
 ).
%  % e)
%  The left column corresponds to
%  the standard real wavenumber, and
%  % f)
%  the right column corresponds to
%  the complex time-causal wavenumber.
 The wavenumbers
 \eqref{eqn:intro_wavenumber} are
 % 1.) standard real wavenumber
 the standard real wavenumber
 (left column) and
 % 2.) complex time-causal wavenumber
 the complex time-causal wavenumber
 (right column).
 % f)
 The numbers of
 % 1.) spatial dimensions
 spatial dimensions are
 % 2.) two
 two
 (top row) and
 % 3.) three
 three
 (bottom row).
 % g)
 All images use
 % 1.) equal axes
 equal axes.
}%
{exp_val_cirs_040_images}
% 
%   The top and bottom rows correspond to the two-dimensional and three-dimensional cases.

%---------------------------------------------------------------------------------------------------------------
% 2.) images [ dynamic apertures ]
%---------------------------------------------------------------------------------------------------------------
% a)
The proposed \name{Laplace}-domain beamforming, as shown in
% 1.) Figure 3
\cref{fig:exp_val_cirs_040_images}, improved
% 2.)
image uniformity,
% 3.)
spatial resolution, and
% 4.)
contrast.
% b)
% lateral widths:
% maximum rel. improvement: 12.04 % [wire: 2, position: (11.80, 6.57) mm]
% LPI (lossy/3d): 498.8382 um +- 148.3387 um [wire 12 @ (1.58, 25.96) mm: 443.7035 um]
% LPI (lossy/2d): 498.8382 um +- 148.3387 um [wire 12: 446.3289 um]
% LPI (lossless/3d): 492.2746 um +- 150.9642 um [wire 12: 446.3289 um]
% LPI (lossless/2d): 489.6491 um +- 152.2769 um [wire 12: 443.7035 um]
% FBP: 533.4000 um +- 154.3050 um [wire 12: 499.1100 um]
% relative improvement: 6.4795 %
% 
% axial widths:
% maximum rel. improvement: 24.65 % [wire: 3, position: (15.12, 11.24) mm]
% LPI (lossy/3d): 405.7799 um +- 46.9850 um (wire 12: 404.3561 um)
% LPI (lossy/2d): 455.6125 um +- 41.2899 um (wire 12: 464.1552 um)
% LPI (lossless/3d): 467.0028 um +- 24.2044 um (wire 12: 478.3931 um)
% LPI (lossless/2d): 465.5790 um +- 18.5093 um (wire 12: 478.3931 um)
% FBP: 455.2950 um +- 45.7200 um (wire 12: 464.8200 um)
% relative improvement: 10.8754 %
% 
% small cyst:
% LPI (lossy/3d): 86.98 %
% LPI (lossy/2d): 84.80 %
% LPI (lossless/3d): 84.19 %
% LPI (lossless/2d): 84.21 %
% FBP: 80.17 %
% 
% large cyst:
% LPI (lossy/3d): 72.57 %
% LPI (lossy/2d): 71.47 %
% LPI (lossless/3d): 65.93 %
% LPI (lossless/2d): 65.13 %
% FBP: 63.95 %
The median lateral and
% 1.) median axial full widths at half maximum (FWHMs)
axial \acp{FWHM}, according to
% 2.) Table I
\cref{tab:metrics}, reduced by
% 3.) ( 533.4000 - 498.8382 ) / 533.4000 * 100 = 6.479527559
up to \qty{\FWHMLateralMedianImprovement}{\percent} and
% 4.) 9.9 %
\qty{\FWHMAxialMedianImprovement}{\percent},
respectively, compared to
% 5.) filtered backpropagation (FBP)
\ac{FBP}.
% c)
The highest reductions of
% 1.a) 12 %
\qty{\FWHMLateralImprovementUB}{\percent} and
% 1.b) 18 %
\qty{\FWHMAxialImprovementUB}{\percent},
respectively, were
% 2.) achieved
achieved for
% 3.) wires
the wires in
% 4.) top right corner
the top right corner.
% d)
The \acp{GCNR} of
% 1.) anechoic regions
the anechoic regions improved by
% 2.) up to 9 %
up to \qty{\gCNRImprovementUB}{\percent}.
% e) time-causal
The complex wavenumber, fixing
% 1.) number
the number of
% 2.) spatial dimensions
spatial dimensions in
% 3.) l2-minimization problem
the optimization
\eqref{eqn:theory_inversion}, improved
% 4.) median axial full widths at half maximum (FWHMs)
the median axial \acp{FWHM} and
% 5.) generalized contrast-to-noise ratios (gCNRs)
the \ac{GCNR} of
% 6.) large anechoic region
the large anechoic region compared to
% 7.) standard real wavenumber
the real wavenumber.
% f)
Three spatial dimensions were
% 1.)
beneficial in combination with
% 2.)
the complex wavenumber.

%%%%%%%%%%%%%%%%%%%%%%%%%%%%%%%%%%%%%%%%%%%%%%%%%%%%%%%%%%%%%%%%%%%%%%%%%%%%%%%%%%%%%%%%%%%%%%%%%%%%%%%%%%%%%%%%
% 6.) conclusion
%%%%%%%%%%%%%%%%%%%%%%%%%%%%%%%%%%%%%%%%%%%%%%%%%%%%%%%%%%%%%%%%%%%%%%%%%%%%%%%%%%%%%%%%%%%%%%%%%%%%%%%%%%%%%%%%
\section{Conclusion}
% \label{sec:conclusion}
%---------------------------------------------------------------------------------------------------------------
% 1.) conclusion
%---------------------------------------------------------------------------------------------------------------
% a)
Attenuation is
% 1.) major effect
a major effect in
% 2.) ultrafast plane-wave imaging
ultrafast plane-wave imaging.
% b)
This effect limits
% 1.) imaging depth
imaging depth and degrades
% 2.) spatial resolution
spatial resolution as
% 3.) depth increases
depth increases.
% c)
The proposed \name{Laplace}-domain beamforming for
% 1.) linear arrays
linear arrays, unlike
% 2.) established Fourier-domain methods
established \name{Fourier}-domain methods, accounts for
% 3.) two underlying mechanisms
two underlying mechanisms, namely
% 4.) three-dimensional diffraction
three-dimensional diffraction and
% 5.) power-law absorption
power-law absorption, while retaining
% 6.) computational efficiency
the computational efficiency.
% d)
The proposed method, compared to
% 1.) filtered backpropagation (FBP)
\ac{FBP}, improves
% 2.) image uniformity
image uniformity,
% 3.) lateral resolution
lateral and
% 4.) axial resolution
axial resolution by
% 5.) up to 12%
up to \qty{\FWHMLateralImprovementUB}{\percent} and
% 6.) up to 25%
\qty{\FWHMAxialImprovementUB}{\percent},
respectively, and
% 7.) contrast
contrast by
% 8.) up to 9%
up to \qty{\gCNRImprovementUB}{\percent}.
% e)
Details of
% 1.) theory
the theory, such as
% 2.) derivation
% the derivation of
% 3.) Born approximation of the scattered acoustic pressure field [ compressibility; Fourier domain; plane-wave decomposition; rigid baffle ]
% \eqref{eqn:theory_p_sc} and
% 4.) computation
the computation of
% 5.) lateral spectrum of the scattered acoustic pressure field [ plane z = 0, discretized, axial coefficient ]
the axial coefficients
\eqref{eqn:theory_lateral_spectrum_discretized_coef_axial}, were
% 6.)
left to
% 7.)
another publication.

% \IEEEtriggeratref{12}
% \IEEEtriggercmd{\enlargethispage{-152mm}}

\bibliographystyle{IEEEtran}

\begin{thebibliography}{1}
\providecommand{\url}[1]{#1}
\csname url@samestyle\endcsname
\providecommand{\newblock}{\relax}
\providecommand{\bibinfo}[2]{#2}
\providecommand{\BIBentrySTDinterwordspacing}{\spaceskip=0pt\relax}
\providecommand{\BIBentryALTinterwordstretchfactor}{4}
\providecommand{\BIBentryALTinterwordspacing}{\spaceskip=\fontdimen2\font plus
\BIBentryALTinterwordstretchfactor\fontdimen3\font minus
  \fontdimen4\font\relax}
\providecommand{\BIBforeignlanguage}[2]{{%
\expandafter\ifx\csname l@#1\endcsname\relax
\typeout{** WARNING: IEEEtran.bst: No hyphenation pattern has been}%
\typeout{** loaded for the language `#1'. Using the pattern for}%
\typeout{** the default language instead.}%
\else
\language=\csname l@#1\endcsname
\fi
#2}}
\providecommand{\BIBdecl}{\relax}
\BIBdecl

\bibitem{proc:SchiffnerAI2012}
M.~F. Schiffner and G.~Schmitz, ``Plane wave pulse-echo ultrasound diffraction
  tomography with a fixed linear transducer array,'' in \emph{Acoust.
  Imaging}.\hskip 1em plus 0.5em minus 0.4em\relax Springer Netherlands, 2012,
  vol.~31, pp. 19--30.

\bibitem{article:ChengITUFFC2006}
J.~Cheng and J.-y. Lu, ``Extended high-frame rate imaging method with
  limited-diffraction beams,'' \emph{IEEE Trans. Ultrason., Ferroelectr., Freq.
  Control}, vol.~53, no.~5, pp. 880--899, May 2006.

\bibitem{article:KruizingaITUFFC2012}
P.~Kruizinga, F.~Mastik, N.~de~Jong, A.~F.~W. van~der Steen, and G.~van Soest,
  ``Plane-wave ultrasound beamforming using a nonuniform fast {F}ourier
  transform,'' \emph{IEEE Trans. Ultrason., Ferroelectr., Freq. Control},
  vol.~59, no.~12, pp. 2684--2691, Dec. 2012.

\bibitem{book:Devaney2012}
A.~J. Devaney, \emph{Mathematical Foundations of Imaging, Tomography and
  Wavefield Inversion}, 1st~ed.\hskip 1em plus 0.5em minus 0.4em\relax
  Cambridge: Cambridge University Press, Jul. 2012.

\bibitem{book:Cobbold2006}
R.~S.~C. Cobbold, \emph{Foundations of Biomedical Ultrasound}.\hskip 1em plus
  0.5em minus 0.4em\relax New York: Oxford University Press, Inc., Sep. 2006.

\bibitem{article:SchiffnerITUFFC2023}
M.~F. Schiffner and G.~Schmitz, ``Frequency-dependent {F}-number suppresses
  grating lobes and improves the lateral resolution in coherent plane-wave
  compounding,'' \emph{IEEE Trans. Ultrason., Ferroelectr., Freq. Control},
  vol.~70, no.~9, pp. 1101--1117, Sep. 2023.

\bibitem{software:LaplaceInversion}
M.~F. Schiffner, ``Laplace-domain beamforming for ultrafast plane-wave
  imaging,'' \url{https://github.com/mschiffn/laplace\_inversion}, 2026.

\bibitem{article:Rodriguez-MolaresITUFFC2020}
A.~Rodriguez-Molares, O.~M.~H. Rindal, J.~D'hooge, S.-E. Måsøy, A.~Austeng,
  M.~A. Lediju~Bell, and H.~Torp, ``The generalized contrast-to-noise ratio:
  {A} formal definition for lesion detectability,'' \emph{IEEE Trans.
  Ultrason., Ferroelectr., Freq. Control}, vol.~67, no.~4, pp. 745--759, Apr.
  2020.

\end{thebibliography}


\end{document}